\documentclass[lettersize,journal]{IEEEtran}
\usepackage{amsmath,amsfonts,amssymb}
\usepackage{mathrsfs}
\usepackage{algorithmic}
\usepackage{algorithm}
\usepackage{array}
\usepackage[caption=false,font=normalsize,labelfont=sf,textfont=sf]{subfig}
\usepackage{textcomp}
\usepackage{stfloats}
\usepackage{url}
\usepackage{verbatim}
\usepackage{graphicx}
\usepackage{cite}
\usepackage{subfig}
\usepackage{caption}

\newtheorem{definition}{Definition}
\newtheorem{proposition}{Proposition}
\newtheorem{remark}{Remark}

\begin{document}

\title{RIS-Enabled Joint Near-Field 3D Localization and Synchronization in SISO Multipath Environments}

\author{Han Yan,~Hua~Chen,~\IEEEmembership{Senior Member,~IEEE},~
        Wei~Liu,~\IEEEmembership{Senior Member,~IEEE},\\
        Songjie Yang,~
        Gang Wang,~\IEEEmembership{Senior Member,~IEEE},~
         and~Chau Yuen,~\IEEEmembership{Fellow,~IEEE}
\thanks{This work was supported by the Zhejiang Provincial Natural Science Foundation of China under Grants LY23F010003 and LR20F010001, by the National Natural Science Foundation of China under Grants 62001256 and 62222109, and the UK Engineering and Physical Sciences Research Council (EPSRC) under Grants EP/V009419/1 and EP/V009419/2.  (\emph{Corresponding author: Hua~Chen}.)}
\thanks{Han Yan, Hua Chen, and Gang Wang are with the Faculty of Electrical Engineering and Computer Science, Ningbo University, Ningbo 315211, China. (e-mail: dkchenhua0714@hotmail.com; wanggang@nbu.edu.cn)}
\thanks{Hua Chen is also with the Zhejiang Key Laboratory of Mobile Network Application Technology,  Ningbo 315211, P. R. China. }
\thanks{Wei Liu is with the School of Electronic Engineering and Computer Science, Queen Mary University of London, London E1 4NS, UK. (e-mail: wliu.eee@gmail.com)} 
\thanks{Songjie Yang is with the National Key Laboratory of Wireless Communications, University of Electronic Science and Technology of China. (e-mail: yangsongjie@std.uestc.edu.cn)}
\thanks{Chau Yuen is with the School of Electrical and Electronics Engineering, Nanyang Technological University, Singapore. (e-mail: chau.yuen@ntu.edu.sg)}
}

\markboth{}
{Shell \MakeLowercase{\textit{et al.}}: A Sample Article Using IEEEtran.cls for IEEE Journals}


\maketitle

\begin{abstract}
Reconfigurable Intelligent Surfaces (RIS) show great promise in the realm of 6th generation (6G) wireless systems, particularly in the areas of localization and communication. Their cost-effectiveness and energy efficiency enable the integration of numerous passive and reflective elements, enabling near-field propagation.
In this paper, we tackle the challenges of RIS-aided 3D localization and synchronization in multipath environments, focusing on the near-field of mmWave systems. Specifically, our approach involves formulating a maximum likelihood (ML) estimation problem for the channel parameters.
To initiate this process, we leverage a combination of canonical polyadic decomposition (CPD) and orthogonal matching pursuit (OMP) to obtain coarse estimates of the time of arrival (ToA) and angle of departure (AoD) under the far-field approximation.
Subsequently, distances are estimated using $l_{1}$-regularization based on a near-field model. Additionally, we introduce a refinement phase employing the spatial alternating generalized expectation maximization (SAGE) algorithm. Finally, a weighted least squares approach is applied to convert channel parameters into position and clock offset estimates.
To extend the estimation algorithm to ultra-large (UL) RIS-assisted localization scenarios, it is further enhanced  to reduce errors associated with far-field approximations, especially in the presence of significant near-field effects, achieved by narrowing the RIS aperture.
Moreover, the Cram$\acute{\text{e}}$r-Rao Bound (CRB) is derived and the RIS phase shifts are optimized to improve the positioning accuracy. Numerical results affirm the efficacy of the proposed estimation algorithm.
\end{abstract}


\section{Introduction}

\IEEEPARstart{I}{ndoorn} positioning plays an important role in the Internet of Things (IoT) and the forthcoming 6G technology, and traditional localization solutions mainly rely technologies primarily relied on Global Position System (GPS) signals or signals from base stations (BSs) \cite{2007,2009,2008,2017}. However, these methods often encounter blind spots due to obstacles, and indoor environments frequently introduce multipath components (MPCs), leading to suboptimal performance for indoor positioning.

In this context, the emergence of reconfigurable intelligent surfaces (RISs) has quickly gained prominence as a promising solution for creating adaptive wireless propagation environments in future communication networks \cite{1JSAC2020,2CST2021,3JSTSP2022,10158690}.
Particularly, when the line-of-sight (LoS) link is obstructed by obstacles, RIS can restore high-precision positioning capabilities by creating a virtual LoS (VLoS) link. Moreover, serving as a reference for synchronized locations, RIS can offer additional geometric measurements. With its large aperture, RIS offers high angular resolution and sufficient distance resolution, enabling positioning of users even in single-input-single-output (SISO) scenarios, even when the LoS path between the BS and the user equipment (UE) is obstructed. Hence, RIS can not only act as a novel means of location reference, but also enhance the positioning accuracy in some challenging scenarios.

Recent studies have highlighted the potential of RIS-aided localization systems in various scenarios
\cite{4JSTSP2022,5arXiv2023,6arXiv2023,7JSTSP2022,8arXiv2023}.
In \cite{4JSTSP2022}, the study delves into the challenge of SISO localization assisted by RIS under spatial-wideband effects and user mobility.
The focus of \cite{5arXiv2023} is on the development of a system capable of simultaneous indoor and outdoor 3D localization, which leverages the unique capabilities of simultaneously transmitting and reflecting RIS (STAR-RIS).
Within \cite{6arXiv2023}, the authors study the application of positioning algorithm to RIS-aided multiple-input-multiple-output (MIMO)
orthogonal frequency division multiplexing (OFDM) systems, considering practical scatterers in the environment.
Moreover, a joint localization and synchronization approach is proposed in \cite{7JSTSP2022}, optimizing the design of active precoding at the base station (BS) and passive phase profiles of the RIS.
Furthermore, \cite{8arXiv2023} addresses the joint RIS calibration and user positioning (JrCUP) problem incorporating an active RIS.

Nonetheless, the aforementioned research primarily assumed that the UE operates in the far field with respect to the RIS.
While this approximation is often convenient, it is not universally applicable, particularly in indoor or in the context of large-scale RIS-assisted positioning scenarios.
In the domain of 3D localization methods involving RIS, there is a conspicuous gap in the literature when it comes to considering the influence of spherical wavefronts in the near-field.
In \cite{9TWC2021} and \cite{10ICC2021}, localization within the near-field range of an RIS acting as a lens is investigated.
Meanwhile, authors in \cite{11TWC2023} grapple with the challenge of RIS-assisted localization under phase-dependent amplitude variations.
Additionally, \cite{10081022} explores the impact of the near-field effect on channel estimation for RIS-enhanced mmWave MIMO communications, followed by a discussion on wideband channel estimation in \cite{2023arXiv230400440Y}. 
Furthermore, in \cite{12JSTSP2023}, researchers delve into the intricacies of localization and channel state information (CSI) estimation in the near-field of a Terahertz (THz) system.
Despite these contributions, to the best of our knowledge, there remains a notable gap in the research concerning RIS-aided near-field 3D localization in multipath environments, a scenario that is especially prevalent in indoor positioning scenarios.

In this paper, we introduce a 3D localization system that utilizes a single antenna BS and a transmitting RIS for simultaneous localization and synchronization, considering the presence of unknown scatterers in the scenario. Two distinct near-field positioning frameworks are presented:
one designed for normal RIS (corresponding to limited near-field effects) and the other for ultra-large (UL) RIS (corresponding to significant near-field effects). In addition, we take into account phase configuration of the RIS to minimize the position error bound (PEB). The primary contributions of this work are as follows:
\begin{itemize}
  \item We focus on the downlink SISO-OFDM configuration within the near-field of a mmWave indoor localization system, incorporating a transmitting RIS. Instead of regarding multipath components as mere sources of noise or interference, our developed algorithms can concurrently estimate the positions of users and scatterers, as well as clock offsets. Additionally, the Cram$\acute{\text{e}}$r-Rao Bound (CRB) is derived for this specific scenario, serving as a benchmark for theoretical performance analysis.
  \item A maximum likelihood (ML) estimation problem is formulated for the channel parameters. To obtain the initial parameter values, we combine tensor decomposition and orthogonal matching pursuit (OMP) \cite{13TIT2011} to obtain preliminary estimates of the time of arrival (ToA) and angle of departure (AoD) at the RIS, using far-field approximation.
Subsequently, we estimate distances employing $l_{1}$-regularization based on a near-field model. To address challenge of high dimensional optimization in ML estimator, a refinement phase is introduced by employing the spatial alternating generalized expectation maximization (SAGE) algorithm \cite{14SAGE1999}. Finally, a weighted least squares (WLS) approach is applied, converting channel parameters to position and clock offset estimates. Simulations illustrate that the far-field approximation can effectively provide an initial solution for ML estimation across most scenarios.
  \item Considering the potential unreliability of coarse estimates in scenarios with a massive RIS due to far-field approximations, the initial algorithm is further modified to ensure robust performance, even in cases of extremely significant near-field effects. Through the use of a tailored RIS phase design, a UL RIS-assisted localization problem is effectively transformed into simultaneous cooperation of multiple sub-RISs. The apertures of each sub-RIS are substantially reduced, resulting in a corresponding reduction in far-field approximation errors. Simulation results show that this approach can achieve superior performance in challenging scenarios.
  \item We introduce a low-complexity method to optimize the phase profile of RIS to further enhance localization accuracy. In situations where there is substantial prior knowledge, the optimized RIS phase design offers a localization accuracy improvement of more than an order of magnitude compared with randomly designed RIS phase configurations. Moreover, it notably bolsters the signal-to-noise ratio (SNR), contributing to the enhancement of communication performance.
\end{itemize}

The remaining part of this paper is structured as follows. In Section \ref{sec:section2}, we introduce the geometry, signal model, and our system assumptions. The CRBs for channel parameters, positions and clock offset are derived in Section \ref{sec:section3}.
Section \ref{sec:section4} presents the overall process and framework of the estimation algorithm. Expanding the scope to UL RIS-assisted localization scenarios, the algorithm discussed in Section \ref{sec:section4}  is extended in Section \ref{sec:section5a}, followed by the optimization of the RIS phase shifts in Section \ref{sec:section5b}.
Numerical results are presented in  Section \ref{sec:section6}, while Section \ref{sec:section7} provides concluding remarks.

\textit{Notations:}
Scalars, vectors, matrices and tensors are denoted by lowercase, boldface lowercase, boldface uppercase, and calligraphic letters, e.g., $x$, $\mathbf{x}$, $\mathbf{X}$, and $\boldsymbol{\mathcal{X}}$, respectively.
The transposition, conjugate, conjugate transpose, pseudoinversion, Hadamard product, outer product and Kronecker product operations are denoted by $(\cdot)^{\mathsf{T}}$, $(\cdot)^{*}$, $(\cdot)^{\mathsf{H}}$, $(\cdot)^{\dagger}$, $\odot$, $\circ$ and $\otimes$, respectively.
The operators $\mathrm{vec}(\mathbf{X})$, $\mathrm{diag}(\mathbf{x})$, $\left\Vert \mathbf{x}\right\Vert _{0}$, $\left\Vert \mathbf{x}\right\Vert _{1}$, $\left\Vert \mathbf{x}\right\Vert$, $\left\Vert \mathbf{X}\right\Vert _{\mathrm{F}}$, $\mathfrak{R}(x)$, $\mathfrak{I}(x)$ represent the vectorization of $\mathbf{X}$, transforms $\mathbf{x}$ to a diagonal matrix, the number of non-zero elements in $\mathbf{x}$, the $l_{1}$ norm of $\mathbf{x}$, the $l_{2}$ norm of $\mathbf{x}$, the Frobenius norm of $\mathbf{X}$, the real part of $x$, the imaginary part of $x$, respectively.
$[\mathbf{x}]_{i}$ denotes the $i$-th element of $\mathbf{x}$ and $[\mathbf{X}]_{i,j}$ is the $(i,j)$-th element of $\mathbf{X}$.
\section{System Model}\label{sec:section2}

\subsection{Geometry Model}
\begin{figure}[htb]
  \centering
  \includegraphics[width=7cm]{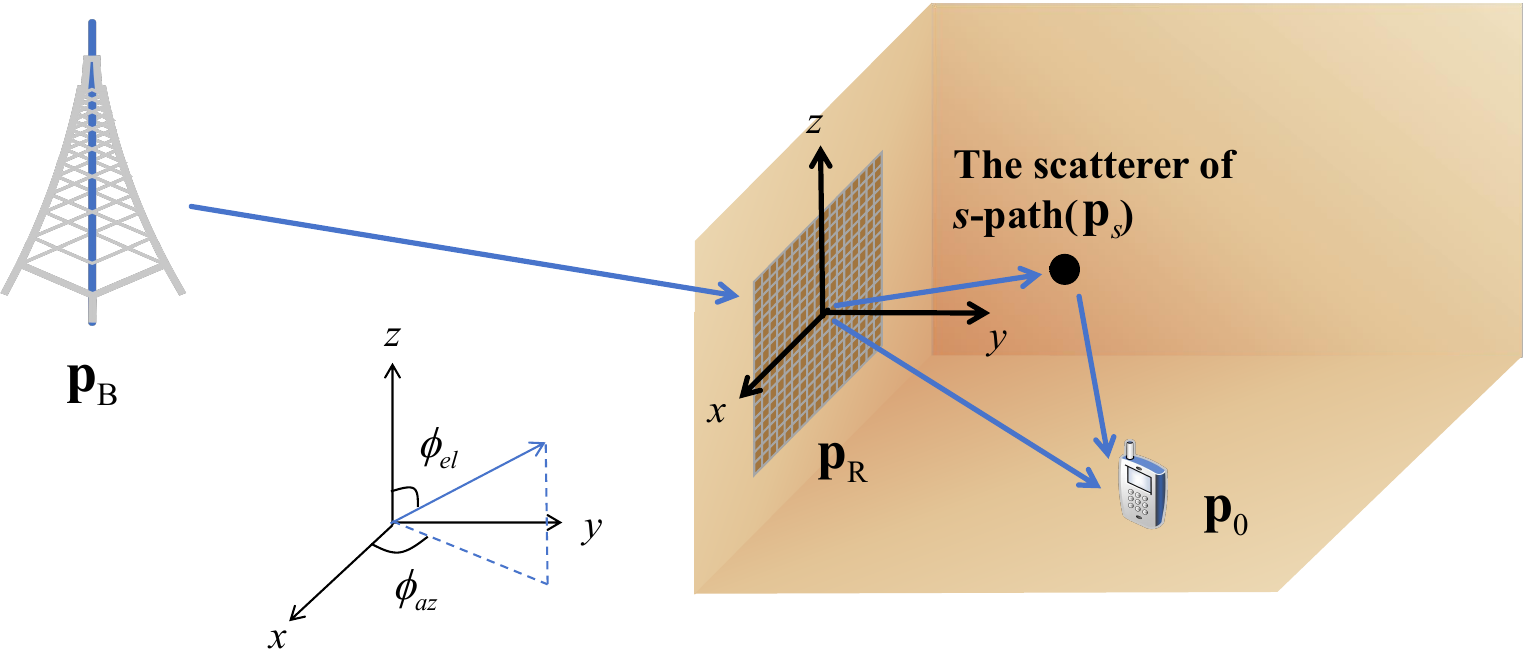}
  \captionsetup{font=small}
  \caption{System model for RIS-assisted localization.}
  \label{fig1}
\end{figure}
As shown in Fig. \ref{fig1}, we consider a downlink RIS-aided mmWave SISO system consisting of a single antenna BS, a transmissive RIS and a single antenna indoor UE. $\mathbf{p}_{\mathrm{B}}\in\mathbb{R}^{3}$ represents the known location of BS, while
$\mathbf{p}_{0}=\left[\mathrm{x}_{0}\text{,}\mathrm{y}_{0},\mathrm{z}_{0}\right]^{\mathsf{T}}$
represents the unknown location of UE. The RIS, comprising $N_{\mathrm{R}}=N_{x}N_{z}$
elements, is placed in parallel to the x-o-z plane with its center located at $\mathbf{p}_{\mathrm{R}}=\left[\mathrm{x}_{\mathrm{R}}\text{,}\mathrm{y}_{\mathrm{R}},\mathrm{z}_{\mathrm{R}}\right]^{\mathsf{T}}$, and $\mathbf{p}_{r}\in\mathbb{R}^{3}$ represents the known location of the $r$-th RIS element for $1\leq r\leq N_{\mathrm{R}}$.
Given the complexity of the indoor environment full of reflected multipath components \cite{15SPM2016,16TWC2023,17JSAC2023}, we consider the presence of $N_{s}$ scatterers with unknown locations. The location of the $s$-th scatterer is denoted by $\mathbf{p}_{s}=\left[\mathrm{x}_{s}\text{,}\mathrm{y}_{s},\mathrm{z}_{s}\right]^{\mathsf{T}},s=1,...,N_{s}$.
The channel parameters in the BS-RIS link can be calculated directly from the coordinates of $\mathbf{p}_{\mathrm{B}}$ and $\mathbf{p}_{\mathrm{R}}$, defined as
\begin{align}
\theta_{el} & =\mathrm{arccos}([\mathbf{p}_{\mathrm{B}}-\mathbf{p}_{\mathrm{R}}]_{3}/\left\Vert \mathbf{p}_{\mathrm{B}}-\mathbf{p}_{\mathrm{R}}\right\Vert ),\label{eq:1}\\
\theta_{az} & =\mathrm{atan}2([\mathbf{p}_{\mathrm{B}}-\mathbf{p}_{\mathrm{R}}]_{2},[\mathbf{p}_{\mathrm{B}}-\mathbf{p}_{\mathrm{R}}]_{1}),
\label{eq:2}\\
d_{\mathrm{B}} & =\left\Vert \mathbf{p}_{\mathrm{B}}-\mathbf{p}_{\mathrm{R}}\right\Vert .\label{eq:3}
\end{align}

Similarly, the unknown channel parameters of the RIS-UE link are defined as
\begin{align}
\phi_{el,s} & =\mathrm{arccos}([\mathbf{p}_{s}-\mathbf{p}_{\mathrm{R}}]_{3}/\left\Vert \mathbf{p}_{s}-\mathbf{p}_{\mathrm{R}}\right\Vert ),\label{eq:4}\\
\phi_{az,s} & =\mathrm{atan}2([\mathbf{p}_{s}-\mathbf{p}_{\mathrm{R}}]_{2},[\mathbf{p}_{s}-\mathbf{p}_{\mathrm{R}}]_{1}),\label{eq:5}\\
d_{s} & =\left\Vert \mathbf{p}_{s}-\mathbf{p}_{\mathrm{R}}\right\Vert ,\label{eq:6}
\end{align}
for $0\leq s\leq N_{s}$. Considering clock offset $\Delta\in\mathbb{R}$ between the BS and the UE , the TOA of all links can be represented as
\begin{align}
\tau_{\mathrm{0}} & =\text{(}\left\Vert \mathbf{p}_{\mathrm{R}}-\mathbf{p}_{\mathrm{B}}\right\Vert +\left\Vert \mathbf{p}_{\mathrm{0}}-\mathbf{p}_{\mathrm{R}}\right\Vert )/c+\Delta,\label{eq:7}\\
\tau_{\mathrm{s}} & =\text{(}\left\Vert \mathbf{p}_{\mathrm{R}}-\mathbf{p}_{\mathrm{B}}\right\Vert +\left\Vert \mathbf{p}_{\mathrm{s}}-\mathbf{p}_{\mathrm{R}}\right\Vert +\left\Vert \mathbf{p}_{\mathrm{0}}-\mathbf{p}_{\mathrm{s}}\right\Vert )/c+\Delta.\label{eq:8}
\end{align}

In this scenario, the RIS is deployed on the side that is closer to the UE \cite{18GLOBECOM2018}. The BS is located in the far-field region of the
RIS, while the indoor environment falls within the Fresnel Near-Field region of the RIS, denoted as \cite{19NF2005}
\begin{equation}
0.62\sqrt{\frac{D^{3}}{\lambda}}\leq d_{s}\leq\frac{2D^{2}}{\lambda},\label{eq:2a}
\end{equation}
where $D$ is the maximum aperture of the RIS and $\lambda$ is the carrier wavelength.

\subsection{Signal Model}

We consider the transmission of $T$ orthogonal
 frequency-division multiplexing (OFDM) pilot symbols with $N$ subcarriers. The frequency of the $n$-th subcarrier is denoted as $f_{n}=f_{c}+n\varDelta f-B/2$,
where $f_{c}$ is the carrier frequency, $\varDelta f$ is the subcarrier spacing, $B=N\varDelta f$ is the bandwidth. The $s_{t}[n]$ is the transmitted signal at the $n$-th subcarrier and the $t$-th transmission with average transmission power $\left|s_{t}[n]\right|=\sqrt{P}$,
where $P$ is the transmit power of the BS. We assume bandwidth $B\ll f_{c}$, in the context of a narrow-band model.

As illustrated in Fig. \ref{fig1}, the RIS-UE link comprises $s$ paths, with the $(s=0)$-th path being the LoS, while the remaining ones correspond to NLoS paths. Subsequently, the channel of the BS-RIS link
$\mathbf{h}_{\mathrm{BR}}[n]\in\mathbb{C}^{N_{\mathrm{R}}}$
and the channel of the RIS-UE link $\mathbf{h}_{\mathrm{RU}}[n]\in\mathbb{C}^{N_{\mathrm{R}}}$
can be respectively modeled as
\begin{align}
\mathbf{h}_{\mathrm{BR}}[n] & =\rho_{\mathrm{BR}}e^{-j2\pi\tau_{\mathrm{BR}}(n-1)\varDelta f}\mathbf{a}\left(\mathbf{p}_{\mathrm{B}}\right),\label{eq:9}\\
\mathbf{h}_{\mathrm{RU}}[n] & =\sum_{s=0}^{N_{s}}\rho_{\mathrm{RU},s}e^{-j2\pi\tau_{\mathrm{RU},s}(n-1)\varDelta f}\mathbf{a}\left(\mathbf{p}_{\mathrm{s}}\right),\label{eq:10}
\end{align}
where $\rho_{\mathrm{BR}}$ and $\tau_{\mathrm{BR}}$ are respectively
the channel gain and TOA of the BS-RIS path, $\rho_{\mathrm{RU},s}$
and $\tau_{\mathrm{RU},s}$ are those of the $s$-th path from the RIS to the UE. $\mathbf{a}\left(\mathbf{p}\right)$
is the near-field RIS steering vector for a given position $\mathbf{p}\in\{\mathbf{p}_{\mathrm{B}},\mathbf{p}_{s}\}$,
defined as
\begin{equation}
[\mathbf{a}(\mathbf{p})]_{r}=\exp(-j2\pi\left(\left\Vert \mathbf{p}-\mathbf{p}_{r}\right\Vert -\left\Vert \mathbf{p}-\mathbf{p}_{\mathrm{R}}\right\Vert \right)/\lambda),\label{eq:11}
\end{equation}
for $r\in\{1,...,N_{\mathrm{R}}\}$. As the distance between the RIS and
the targets (UE, scatters) becomes significantly larger compared to
the size of the RIS, the near-field steering vector described in (\ref{eq:11})
converges to its conventional far-field counterpart \cite{20TSP2019}.

The received signal at the UE for the $n$-th subcarrier and the $t$-th
transmission can be written as
\begin{align}
y_{t}[n] & =\mathbf{h}_{\mathrm{BR}}^{\mathsf{T}}[n]\mathrm{diag}(\mathbf{w}_{t})\mathbf{h}_{\mathrm{RU}}[n]s_{t}[n]+z_{t}[n]\nonumber \\
 & =\sum_{s=0}^{N_{s}}\rho_{s}e^{-j2\pi\tau_{s}(n-1)\varDelta f}\mathbf{b}^{\mathsf{T}}\left(\mathbf{p}_{s}\right)\mathbf{w}_{t}s_{t}[n]+z_{t}[n],\label{eq:12}
\end{align}
where $\rho_{s}\triangleq\rho_{BR}\rho_{RU,s}$, $\tau_{s}\triangleq\tau_{\mathrm{BR}}+\tau_{RU,s}$,
$\mathbf{b}(\mathbf{p})=\mathbf{a}\left(\mathbf{p}\right)\odot\mathbf{a}\left(\mathbf{p}_{\mathrm{B}}\right)$,
$\mathbf{w}_{t}=\left[w_{t,1}\ldots w_{t,N_{\mathrm{R}}}\right]^{\mathsf{T}}$
is the RIS phase shifts at the transmission $t$, and ${z}_{t}[n]$
is the zero-mean additive Gaussian noise with variance $N_{0}$. For
simplicity, assume that all the transmitted pilot symbols are equal to $\sqrt{P}$. By defining $\mathbf{c}^{(N_{m})}(\omega)=[1,e^{j\omega},...,e^{j(N_{m}-1)\omega}]^{\mathsf{T}}\in\mathbb{C}^{N_{m}},$
the received signal can be rewritten in an $N\times T$ matrix
\begin{equation}
\mathbf{Y}=\sqrt{P}\sum_{s=0}^{N_{s}}\rho_{s}\mathbf{c}^{(N)}(\omega_{s}^{(1)})\mathbf{b}^{\mathsf{T}}\left(\mathbf{p}_{s}\right)\mathrm{\mathbf{W}}+\mathbf{Z},\label{eq:13}
\end{equation}
where $\mathbf{W}=\left[\mathbf{w}_{1},\mathbf{w}_{2},\ldots,\mathbf{w}_{T}\right]\in\mathbb{C}^{N_{\mathrm{R}}\times T}$, $[\mathbf{Z}]_{n,t}=z_{t}[n]$, and
\begin{equation}
\omega_{s}^{(1)}=-2\pi\tau_{s}\varDelta f.\label{eq:14}
\end{equation}

\section{CRB Analyses}\label{sec:section3}

In this section, we establish the Fisher Information Matrix (FIM) and the CRB for the joint localization and synchronization estimation task, which will serve as a reference point to gauge the accuracy of the proposed estimation algorithms.

\subsection{CRB for Channel Parameter Estimation}\label{sec:section3a}

We define a vector consisting of the unknown channel parameters as
$\boldsymbol{\eta}=[\boldsymbol{\eta}_{0}^{\mathsf{T}},...,\boldsymbol{\eta}_{s}^{\mathsf{T}},...,\boldsymbol{\eta}_{N_{s}}^{\mathsf{T}}]^{\mathsf{T}}\in\mathbb{R}^{6(N_{s}+1)}$
with $\boldsymbol{\eta}_{s}=[\mathfrak{R}(\rho_{s}),\mathfrak{I}(\rho_{s}),\phi_{el,s},\phi_{az,s},d_{s},\tau_{s}]$$^{\mathsf{T}}$.
Subsequently, the channel parameter CRB can be obtained as
$\mathbf{F}(\boldsymbol{\eta})^{-1}\in\mathbb{R}^{6(N_{s}+1)\times6(N_{s}+1)}$,
with the FIM of the channel parameter vector defined as \cite{21CRB1993}:
\begin{equation}
\mathbf{F}(\boldsymbol{\eta})=\frac{2}{\sigma^{2}}\sum_{t=1}^{T}\sum_{n=1}^{N}\mathfrak{R}\left\{ \left(\frac{\partial\mu_{t}[n]}{\partial\boldsymbol{\eta}}\right)^{\mathrm{\mathsf{H}}}\left(\frac{\partial\mu_{t}[n]}{\partial\boldsymbol{\eta}}\right)\right\}
.\label{eq:15}
\end{equation}

Here, the observation $\mu_{t}[n]$ is defined as the noise-free received signal observation:
\begin{equation}
\mu_{t}[n]=\sqrt{P}\sum_{s=0}^{N_{s}}\rho_{s}e^{-j2\pi\tau_{s}(n-1)\varDelta f}\mathbf{b}^{\mathsf{T}}\left(\mathbf{p}_{s}\right)\mathrm{\mathbf{w}_{t}}.\label{eq:16}
\end{equation}

For detailed derivations of \eqref{eq:15}, please refer to Appendix A.

\subsection{CRB for 3D Positioning}\label{sec:section3b}

In order to determine the FIM in the position space,
we perform a variable transformation from channel parameters $\boldsymbol{\eta}$ to
$\boldsymbol{\eta}_\mathrm{p}=[\mathbf{p}^{\mathsf{T}},\Delta,\mathfrak{R}(\boldsymbol{\rho}^{\mathsf{T}}),\mathfrak{I}(\boldsymbol{\rho}^{\mathsf{T}})]^{\mathsf{T}}\in\mathbb{R}^{5N_{s}+6}$,
where $\mathbf{p}=[\mathbf{p}_{0}^{\mathsf{T}},...,\mathbf{p}_{s}^{\mathsf{T}},...,\mathbf{p}_{N_{s}}^{\mathsf{T}}]^{\mathsf{T}}$, and $\boldsymbol{\rho}=[\rho_{0},...,\rho_{s},...,\rho_{N_{s}}]^{\mathsf{T}}$.
The FIM for $\boldsymbol{\eta}_\mathrm{p}$ can be derived using the chain rule \cite{21CRB1993}:
\begin{equation}
\mathbf{F}(\boldsymbol{\eta}_\mathrm{p})=\mathbf{J}\mathbf{F}(\boldsymbol{\eta})\mathbf{J}^{\mathsf{T}},\label{eq:17}
\end{equation}
where
$\mathbf{J}\in\mathbb{R}^{(5N_{s}+6)\times6(N_{s}+1)}$ is the Jacobian matrix defined as
$\mathbf{J}\triangleq\partial\boldsymbol{\eta}^{\mathsf{T}}/\partial\boldsymbol{\eta}_\mathrm{p}$
and it is explicitly provided in Appendix B.
The position error bound (PEB) can be calculated as
\begin{equation}
\mathrm{PEB}=\sqrt{\mathrm{tr}\{[\mathbf{F}(\boldsymbol{\eta}_\mathrm{p})^{-1}]_{1:3,1:3}\}}.\label{eq:18}
\end{equation}

Similarly, the clock offset error bound (CEB) is obtained as
\begin{equation}
\mathrm{CEB}=\sqrt{[\mathbf{F}(\boldsymbol{\eta}_\mathrm{p})^{-1}]_{(3N_{s}+4),(3N_{s}+4)}}.\label{eq:19}
\end{equation}

\section{Estimation Algorithm}\label{sec:section4}

In this section, we first approximate the representation of the received signal in (\ref{eq:13}) through the use of tensor notation. Then, we present an estimator designed to yield a preliminary estimation of the channel parameters. Subsequently, we apply a refinement process to enhance the accuracy of all parameter estimates, employing the SAGE algorithm. This refinement approach is consistent with established practices within the field of localization and is frequently employed in literature \cite{22TWC2018}. Lastly, we leverage the EXIP \cite{23EXIP1989} to estimate the UE position and clock offset.

\subsection{Tensor Representation}\label{sec:section4a}

To convert the matrix in (\ref{eq:13}) into a tensor, we approximate
the near-field steering vector by its far-field counterpart \cite{24ICC2021}
\begin{equation}
\mathbf{a}(\mathbf{p})\approx\mathbf{a}(\varphi_{el},\varphi_{az})\triangleq e^{j(\beta_{\varphi,x}+\beta_{\varphi,z})}\mathbf{c}^{(N_{x})}(\varphi_{x})\otimes\mathbf{c}^{(N_{z})}(\varphi_{z}),\label{eq:20}
\end{equation}
where $\beta_{\varphi,x}=-\pi \mathrm{sin}\varphi_{el}\mathrm{cos}\varphi_{az}(N_{x}-1)d/\lambda$,
$\beta_{\varphi,z}=-\pi \mathrm{cos}\varphi_{el}(N_{z}-1)d/\lambda$,
$\varphi_{x}=2\pi \mathrm{sin}\varphi_{el}\mathrm{cos}\varphi_{az}d/\lambda$, and
$\varphi_{z}=2\pi \mathrm{cos}\varphi_{el}d/\lambda$.

After some simplifications, the received signal can be approximately expressed as
\begin{equation}
\mathbf{Y}=\sum_{s=0}^{N_{s}}\tilde{\rho}_{s}\mathbf{c}^{(N)}(\omega_{s}^{(1)})\circ\mathbf{W}^{\mathsf{T}}[\mathbf{c}^{(N_{x})}(\omega_{s}^{(2)})\otimes\mathbf{c}^{(N_{z})}(\omega_{s}^{(3)})]+\mathbf{Z}\label{eq:21}
\end{equation}
where $\tilde{\rho}_{s}=\sqrt{P}\rho_{s}e^{j(\beta_{\theta,x}+\beta_{\theta,z}+\beta_{\phi,x}+\beta_{\phi,z})}$, and
\begin{align}
\omega_{s}^{(2)} & =2\pi(\mathrm{sin}\theta_{el}\mathrm{cos}\theta_{az}+\mathrm{sin}\phi_{el,s}\mathrm{cos}\phi_{az,s})d/\lambda,\label{eq:22}\\
\omega_{s}^{(3)} & =2\pi(\mathrm{cos}\theta_{el}+\mathrm{cos}\phi_{el,s})d/\lambda.\label{eq:23}
\end{align}

Similar to \cite{8arXiv2023}, we construct a total RIS profile matrix $\mathbf{W}$ as follows
\begin{equation}
\mathbf{W}=\mathbf{T}_{1}\otimes\mathbf{T}_{2}\in\mathbb{C}^{N_{x}N_{z}\times T},\label{eq:24}
\end{equation}
where $\mathbf{T}_{1}\in\mathbb{C}^{N_{x}\times T_{1}}$, $\mathbf{T}_{2}\in\mathbb{C}^{N_{z}\times T_{2}}$,
and $T=T_{1}T_{2}$. It can be further obtained that
\begin{equation}
\mathbf{Y}=\sum_{s=0}^{N_{s}}\tilde{\rho}_{s}\mathbf{c}^{(N)}(\omega_{s}^{(1)})\circ[\mathbf{T}_{1}^{\mathsf{T}}\mathbf{c}^{(N_{x})}(\omega_{s}^{(2)})\otimes\mathbf{T}_{2}^{\mathsf{T}}\mathbf{c}^{(N_{z})}(\omega_{s}^{(3)})]+\mathbf{Z}.\label{eq:25}
\end{equation}

Then the received signal can be represented as a three-order tensor
$\boldsymbol{\mathcal{Y}}\in\mathbb{C}^{N\times T_{1}\times T_{2}}$
\begin{align}
\boldsymbol{\mathscr{\mathcal{Y}}} & =\sum_{s=0}^{N_{s}}\tilde{\rho}_{s}\mathbf{c}^{(N)}(\omega_{s}^{(1)})\circ\mathbf{T}_{1}^{\mathsf{T}}\mathbf{c}^{(N_{x})}(\omega_{s}^{(2)})\circ\mathbf{T}_{2}^{\mathsf{T}}\mathbf{c}^{(N_{z})}(\omega_{s}^{(3)})+\boldsymbol{\mathcal{Z}}\nonumber \\
 & \triangleq\sum_{s=0}^{N_{s}}\tilde{\rho}_{s}\mathbf{r}_{1}(\omega_{s}^{(1)})\circ\mathbf{r}_{2}(\omega_{s}^{(2)})\circ\mathbf{r}_{3}(\omega_{s}^{(3)})+\boldsymbol{\mathcal{Z}}.\label{eq:26}
\end{align}
where $\mathbf{r}_{1}(\omega_{s}^{(1)})\triangleq\mathbf{c}^{(N)}(\omega_{s}^{(1)})$,
$\mathbf{r}_{2}(\omega_{s}^{(2)})\triangleq\mathbf{T}_{1}^{\mathsf{T}}\mathbf{c}^{(N_{x})}(\omega_{s}^{(2)})$, and
$\mathbf{r}_{3}(\omega_{s}^{(3)})\triangleq\mathbf{T}_{2}^{\mathsf{T}}\mathbf{c}^{(N_{z})}(\omega_{s}^{(3)})$.

\subsection{Estimation of Channel Parameters}\label{sec:section4b}
\subsubsection{Coarse Estimation of TOAs and AODs}
The overall approach follows the concept of OMP, obtaining the estimates for one path at each iteration.
Concurrently, we utilize canonical polyadic decomposition (CPD) to separate the signal components for each path, and converting the 3D parameter estimation problem involving TOA and AOD parameters in
(\ref{eq:14}), (\ref{eq:22}), (\ref{eq:23}) into three separate 1D search problems, thereby reducing the overall complexity.
\begin{definition}[CP decomposition \cite{25CPD2017,CPD2009,CPD2000}]
The CPD, also known as PARAFAC, decomposes tensor data $\boldsymbol{\mathscr{\mathcal{X}}} \in \mathbb{R}^{I_{1} \times \cdots \times I_{N}}$
into a sum of R rank-1 tensors:
\begin{equation}
\boldsymbol{\mathscr{\mathcal{X}}} = \sum_{r=1}^{R} \underbrace{\mathbf{v}_{r}^{(1)} \circ \cdots \circ \mathbf{v}_{r}^{(N)}}_{\text{rank-1 tensor}}.\label{eq:27}
\end{equation}

\end{definition}

In the $s$-th iteration, a rank-1 CPD is used to separate
the signal component corresponding to the $s$-th path:
\begin{equation}
\boldsymbol{\mathscr{\mathcal{Y}}}_{s}\approx\mathbf{u}_{s}^{(1)}\circ\mathbf{u}_{s}^{(2)}\circ\mathbf{u}_{s}^{(3)},\label{eq:28}
\end{equation}
where $\mathbf{u}_{s}^{(n)}(n=1,2,3)$ is the factor vector along
the $n$-th mode with the expression of
\begin{equation}
\mathbf{u}_{s}^{(n)}=\alpha_{s}^{(n)}\mathbf{r}_{n}(\omega_{s}^{(n)}),\label{eq:29}
\end{equation}
where $\alpha_{s}^{(n)}\in\mathbb{C}$.
The estimation of $\alpha_{s}^{(n)}$ and $\omega_{s}^{(n)}$ in (\ref{eq:29})
can be formulated as
\begin{equation}
[\hat{\alpha}_{s}^{(n)},\hat{\omega}_{s}^{(n)}]=\underset{\alpha_{s}^{(n)},\omega_{s}^{(n)}}{\arg\min}\left\Vert \mathbf{u}_{s}^{(n)}-\alpha_{s}^{(n)}\mathbf{r}_{n}(\omega_{s}^{(n)})\right\Vert.\label{eq:30}
\end{equation}

Here, $\alpha_{s}^{(n)}$
as a function of $\omega_{s}^{(n)}$
can be derived in closed form as follows:
\begin{equation}
\hat{\alpha}_{s}^{(n)}=\mathbf{r}_{n}(\omega_{s}^{(n)})^{\dagger}\mathbf{u}_{s}^{(n)}.\label{eq:31}
\end{equation}

Therefore, coarse estimates of TOA and AOD can be obtained by
solving the following three 1D search problems:
\begin{equation}
\hat{\omega}_{s}^{(n)}=\arg\min_{\omega_{s}^{(n)}}\left\Vert \mathbf{u}_{s}^{(n)}-\hat{\alpha}_{s}^{(n)}\mathbf{r}_{n}(\omega_{s}^{(n)})\right\Vert. \label{eq:32}
\end{equation}

To remove the correlated components of the signal in the $s$-th iteration,
and obtain the updated residual, we subtract the projection of the
above signal using the following procedure:
\begin{align}
\mathbf{A}_{s} & =\mathbf{r}_{1}(\omega_{s}^{(1)})\circ[\mathbf{r}_{2}(\omega_{s}^{(2)})\otimes\mathbf{r}_{3}(\omega_{s}^{(3)})],\label{eq:33}\\
\mathbf{y}_{s+1} & =\mathbf{y}_{s}-\mathrm{vec}(\mathbf{A}_{s})\mathrm{vec}(\mathbf{A}_{s})^{\dagger}\mathbf{y}_{s}.\label{eq:34}
\end{align}

Here, $\mathbf{y}_{s}=\mathrm{vec}(\boldsymbol{\mathscr{\mathcal{Y}}}_{s})$.
Using the obtained updated residual $\mathbf{y}_{s+1}\in\mathbb{C}^{NT_{1}T_{2}}$,
we can reconstruct the tensor $\boldsymbol{\mathscr{\mathcal{Y}}}_{s+1}$
and proceed to the next iteration. The proposed algorithm,
called CPD-OMP, is summarized in Algorithm 1.

\begin{algorithm}[H]
\caption{CPD-OMP to estimate TOA and AOD}\label{alg:alg1}
\renewcommand{\algorithmicrequire}{\textbf{Input:}}
\renewcommand{\algorithmicensure}{\textbf{Output:}}
\begin{algorithmic}[1]
\REQUIRE Recieved signal matrix $\mathbf{Y}$, RIS profile matrix $\mathbf{W}$.
\ENSURE $\{\hat{\tau}_{s}\}_{s=0}^{N_{s}}$, $\{\hat{\phi}_{el,s}\}_{s=0}^{N_{s}}$, $\{\hat{\phi}_{az,s}\}_{s=0}^{N_{s}}$.
\STATE \textbf{Initialization:} Set $\mathbf{y}_{0}=\mathrm{vec}(\mathbf{Y})$ and $s=0$.
\WHILE{$s\leq N_{s}$}
\STATE Construct the tensor $\boldsymbol{\mathscr{\mathcal{Y}}}_{s}$ from $\mathbf{y}_{s}$,
and perform a rank-1 CPD to obtain
$\mathbf{\boldsymbol{\mathrm{\mathit{u}}}_{\mathit{s}}^{\text{(1)}}}$,
$\mathbf{\boldsymbol{\mathrm{\mathit{u}}}_{\mathit{s}}^{\text{(2)}}}$, and
$\mathbf{\boldsymbol{\mathrm{\mathit{u}}}_{\mathit{s}}^{\text{(3)}}}$.
\STATE Estimate ${\omega}_{s}^{(1)}$, ${\omega}_{s}^{(2)}$, and ${\omega}_{s}^{(3)}$ using
\eqref{eq:32}.
\STATE Obtain $\hat{\tau}_{s}$, $\hat{\phi}_{el,s}$ and $\hat{\phi}_{az,s}$
using \eqref{eq:14}, \eqref{eq:22} and \eqref{eq:23}.
\STATE Update residual using \eqref{eq:34}.
\STATE Update $s=s+1$.
\ENDWHILE
\end{algorithmic}
\label{alg1}
\end{algorithm}

\begin{remark}
For the case where the number of scatterers $N_{s}$ is unknown, the decision to continue iterations can be made by comparing the magnitude of the residual fitting error $\left\Vert \mathbf{y}_{s+1}-\mathbf{y}_{s}\right\Vert ^{2}$ with a threshold $\delta$. The value for $\delta$ can be obtained according to \cite{22TWC2018,JSTSP2016}.
\end{remark}

\subsubsection{Coarse Estimation of Distances and Channel Gains}

Once we have obtained the coarse estimates of TOA and AOD, we can
rewrite the received signal as follows:
\begin{equation}
\mathbf{Y}=\sqrt{P}\sum_{s=0}^{N_{s}}\rho_{s}\mathbf{r}_{1}(\hat{\omega}_{s}^{(1)})\mathbf{b}^{\mathsf{T}}\left(\mathbf{p}(d_{s},\hat{\phi}_{el,s},\hat{\phi}_{az,s})\right)\mathrm{\mathbf{W}}+\mathbf{Z},\label{eq:26}
\end{equation}
where $\mathbf{p}(d,\varphi_{el},\varphi_{az})=\mathbf{p}_{\mathrm{R}}+d\mathbf{k}(\varphi_{el},\varphi_{az})$,
and
\begin{equation}
\mathbf{k}(\varphi_{el},\varphi_{az})\triangleq[\mathrm{sin}\varphi_{el}\mathrm{cos}\varphi_{az},\mathrm{sin}\varphi_{el}\mathrm{sin}\varphi_{az},\mathrm{cos}\varphi_{el}]^{\mathsf{T}}.
\end{equation}
Due to the sparse characteristics of received signals in the spatial
domain, the overcomplete dictionary $\mathbf{D}_{s}\in\mathbb{C}^{NT\times M}$
corresponding to the $s$-th path is first constructed with $M$ being the number
of grid samples as
\begin{equation}
\mathbf{D}_{s}=\left[\mathbf{d}_{s}\left(d_{1}\right),\ldots,\mathbf{d}_{s}\left(d_{m}\right),\ldots,\mathbf{d}_{s}\left(d_{M}\right)\right], \label{eq:27}
\end{equation}
where $\mathbf{d}_{s}\left(d_{m}\right)=\mathrm{vec}(\mathbf{r}_{1}(\hat{\omega}_{s}^{(1)})\mathbf{b}^{\mathsf{T}}\left(\mathbf{p}(d_{m},\hat{\phi}_{el,s},\hat{\phi}_{az,s})\right)\mathrm{\mathbf{W}})$, and
$\{d_{m}\}_{m=1}^{M}$is the sampling grid set that covers potential
distance values.

With the aid of the overcomplete dictionary $\mathbf{D}_{s}$, we
can formulate the vectorization of the received signal vector, namely
$\mathbf{y}=\mathrm{vec}(\mathbf{Y})\in\mathbb{C}^{NT}$, into an expression
of sparse representation as follows:
\begin{equation}
\mathbf{y}=\sum_{s=0}^{N_{s}}\mathbf{D}_{s}\boldsymbol{\zeta}_{s}+\mathbf{z},\label{eq:37}
\end{equation}
where $\boldsymbol{\zeta}_{s}\in\mathbb{C}^{M}$ denotes the sparse
vector, and $\mathbf{z}$ represents the noise component.

It can be seen that the above overcomplete representation has transformed
the distance estimation problem into one for estimating parameterized
vectors $\boldsymbol{\zeta}_{s}$ by solving the following optimization
problem:
\begin{subequations}
\begin{align}
    \underset{\boldsymbol{\zeta}_{s}}{\text{min}}\quad & \left\Vert \mathbf{y} - \sum_{s=0}^{N_{s}} \mathbf{D}_{s} \boldsymbol{\zeta}_{s} \right\Vert, \label{eq:38a} \\
    \text{s.t.} \quad & \left\Vert \boldsymbol{\zeta}_{s} \right\Vert_{0} = 1 .\label{eq:38b}
\end{align}
\end{subequations}

Due to the non-convexity of the problem (\ref{eq:38a}), we
can relax the $l_{0}$ norm to the $l_{1}$ norm, which leads to an
optimization problem in the form of LASSO:
\begin{equation}
\hat{\boldsymbol{\zeta}}=\underset{\boldsymbol{\zeta}}{\arg\min}\left\Vert \mathbf{y}-\mathbf{D}\boldsymbol{\zeta}\right\Vert +\xi\|\boldsymbol{\zeta}\|_{1},\label{eq:39}
\end{equation}
where $\xi$ is a regularization parameter, $\boldsymbol{\zeta}=\left[\mathbf{\boldsymbol{\zeta}}_{0}^{\mathsf{T}},\ldots,\mathbf{\boldsymbol{\zeta}}_{N_{s}}^{\mathsf{T}}\right]^{\mathsf{T}}$, and
$\mathbf{D}=\left[\mathbf{D}_{0},\ldots,\mathbf{D}_{N_{s}}\right]$.
The problem in (\ref{eq:39}) can be solved using existing convex solvers \cite{CVX}.
After estimating $\boldsymbol{\zeta}$ , the distance parameters for each path can be determined separately by plotting it on the predefined search grid of potential distances.

With estimates of the distance parameter, the received signal
can be represented as
\begin{align}
\mathbf{Y} & =\sqrt{P}\mathbf{C}(\hat{\boldsymbol{\omega}}^{(1)})\mathrm{diag}(\mathbf{\boldsymbol{\rho}})\mathbf{B}^{\mathsf{T}}\left(\mathbf{p}(\hat{\mathbf{d}},\hat{\boldsymbol{\phi}}_{el},\hat{\boldsymbol{\phi}}_{az})\right)\mathrm{\mathbf{W}}+\mathbf{Z}\nonumber \\
 & \triangleq\mathbf{C}(\hat{\boldsymbol{\omega}}^{(1)})\mathrm{diag}(\mathbf{\boldsymbol{\rho}})\mathbf{Q}+\mathbf{Z},\label{eq:40}
\end{align}
where $\mathbf{C}(\hat{\boldsymbol{\omega}}^{(1)})=[\mathbf{c}^{(N)}(\hat{\omega}_{0}^{(1)}),...,\mathbf{c}^{(N)}(\hat{\omega}_{N_{s}}^{(1)})]\in\mathbb{C}^{N\times N_{s}}$,
$\boldsymbol{\rho}=[\rho_{0},...,\rho_{N_{s}}]$, $\mathbf{B}\left(\mathbf{p}(\cdot)\right)=[\mathbf{b}\left(\mathbf{p}_{0}(\cdot)\right),...,\mathbf{b}\left(\mathbf{p}_{N_{s}}(\cdot)\right)]\in\mathbb{C}^{N_{\mathrm{R}}\times N_{s}}$,
and $\mathbf{Q}\triangleq\sqrt{P}\mathbf{B}^{\mathsf{T}}\left(\mathbf{p}(\hat{\mathbf{d}},\hat{\boldsymbol{\phi}}_{el},\hat{\boldsymbol{\phi}}_{az})\right)\mathrm{\mathbf{W}}$.

Further, one can obtain
\begin{equation}
\mathbf{YQ^{\dagger}}=\mathbf{C}(\hat{\boldsymbol{\omega}}^{(1)})\mathrm{diag}(\mathbf{\boldsymbol{\rho}}).\label{eq:41}
\end{equation}

Then, the complex channel gain can be estimated by
\begin{equation}
\hat{\rho}_{s}=[\mathbf{C}(\hat{\boldsymbol{\omega}}^{(1)})]_{:,s}^{\dagger}[\mathbf{Y}\mathbf{Q}^{\dagger}]_{:,s}.\label{eq:42}
\end{equation}

\subsubsection{Refinement of Channel Parameters}

Due to the coarse estimation of AOD using the far-field approximation, it has a certain impact on the estimation performance and further affects the accuracy of distance parameter estimation. In addition, the precision of parameter 1D search is also influenced by the grid size. Therefore, we consider utilizing ML estimator to jointly refine all these channel parameters. Firstly, based on the signal model, we can construct the following maximum likelihood estimator:
\begin{equation}
\hat{\boldsymbol{\eta}}_{\mathrm{ML}}=\arg\min_{\boldsymbol{\mathbf{\eta}}}\parallel\mathbf{Y}-\mathbf{\Gamma}(\boldsymbol{\mathbf{\eta}})\parallel_{\mathrm{F}},\label{eq:43}
\end{equation}
where $\mathbf{\Gamma}(\boldsymbol{\mathbf{\eta}})=\sqrt{P}\sum_{s=0}^{N_{s}}\rho_{s}\mathbf{c}^{(N)}(\omega_{s}^{(1)})\mathbf{b}^{\mathsf{T}}\left(\mathbf{p}_{s}\right)\mathrm{\mathbf{W}}$.
Unfortunately, solving optimization problem (\ref{eq:43}) entails high dimensional nonlinear optimization for
$\boldsymbol{\eta}\in\mathbb{R}^{6(N_{s}+1)}$,
resulting in significant computational complexity.
Therefore, the space alternating generalized
expectation (SAGE) algorithm is utilized by representing the incomplete data space
$\mathbf{Y}$ as a superposition of $N_{s}+1$ complete data spaces
$\mathbf{Y}_{s}$, as follows
\begin{equation}
\mathbf{Y}=\sum_{s=0}^{N_{s}}\underbrace{\mathbf{\Gamma}_{s}(\boldsymbol{\mathbf{\eta}}_{s})+\mathbf{Z}_{s}}_{\mathbf{Y}_{s}},\label{eq:44}
\end{equation}
where $\mathbf{\Gamma}_{s}(\boldsymbol{\mathbf{\eta}}_{s})=\sqrt{P}\rho_{s}\mathbf{r}_{1}(\omega_{s}^{(1)})\mathbf{b}^{\mathsf{T}}\left(\mathbf{p}(d_{s},\phi_{el,s},\phi_{az,s})\right)\mathrm{\mathbf{W}}$.
We can estimate $\mathbf{Y}_{s}$ based on the observation $\mathbf{Y}$
of the incomplete data and the previous estimation of $\boldsymbol{\eta}$.
At the $(i+1)$-th iteration, we estimate the received signal of the $s$-th path as follows:
\begin{equation}
\hat{\mathbf{Y}}_{s}^{i+1}=\mathbb{E}\left(\mathbf{Y}_{s}\mid\mathbf{Y},\hat{\boldsymbol{\eta}}^{i}\right),\label{eq:45}
\end{equation}
Through (\ref{eq:45}), have \cite{TASSP1988,6arXiv2023}
\begin{equation}
\hat{\mathbf{Y}}_{s}^{i+1}=\mathbf{Y}-\sum_{s'=0}^{s-1}\mathbf{\Gamma}_{s}(\hat{\boldsymbol{\eta}}_{s'}^{i+1})-\sum_{s''=s+1}^{N_{s}}\mathbf{\Gamma}_{s}(\hat{\boldsymbol{\eta}}_{s''}^{i}),\label{eq:46}
\end{equation}
and consequently, the channel parameters of the $s$-th path are refined by solving the optimization problem given by
\begin{equation}
\hat{\boldsymbol{\eta}}_{s}^{i+1}=\arg\min_{\boldsymbol{\mathbf{\eta}}_{s}}\left\Vert \hat{\mathbf{Y}}_{s}^{i+1}-\mathbf{\Gamma}_{s}(\boldsymbol{\mathbf{\eta}}_{s})\right\Vert_{\mathrm{F}}.\label{eq:47}
\end{equation}

We can employ the Nelder-Mead algorithm \cite{Nelder-Mead} to solve (\ref{eq:47}) using results of the coarse estimation as initial values. The Nelder-Mead method is renowned for its rapid convergence and does not depend on derivative information. The comprehensive SAGE algorithm for refining the channel parameters in $\boldsymbol{\mathbf{\eta}}$
is outlined in Algorithm 2.

\begin{algorithm}[H]
\caption{Refine channel parameters using SAGE}\label{alg:alg2}
\renewcommand{\algorithmicrequire}{\textbf{Input:}}
\renewcommand{\algorithmicensure}{\textbf{Output:}}
\begin{algorithmic}[1]
\REQUIRE Recieved signal matrix $\mathbf{Y}$, coarse estimates of channel parameters $\hat{\boldsymbol{\eta}}^{c}$,
convergence threshold $\epsilon$ and maximum number of iterations
$I$.
\ENSURE Refined channel parameters $\hat{\boldsymbol{\eta}}$.
\STATE \textbf{Initialization:} Set $\hat{\boldsymbol{\eta}}^{0}=\hat{\boldsymbol{\eta}}^{c}$ and $i=0$.
\WHILE{$i\leq I$}
\STATE Set $s=0$, $i=i+1$.
\WHILE{$s\leq N_{s}$}
\STATE Estimate $\hat{\mathbf{Y}}_{s}^{i}$ by using \eqref{eq:46}.
\STATE Estimate $\hat{\boldsymbol{\eta}}_{s}^{i}$ by using \eqref{eq:47}.
\STATE Update\\
$\hat{\boldsymbol{\eta}}^{i}=[(\hat{\boldsymbol{\eta}}_{0}^{i})^{\mathsf{T}},...,(\hat{\boldsymbol{\eta}}_{s}^{i})^{\mathsf{T}},(\hat{\boldsymbol{\eta}}_{s+1}^{i-1})^{\mathsf{T}},...,(\hat{\boldsymbol{\eta}}_{N_{s}}^{i-1})^{\mathsf{T}}]^{\mathsf{T}}$.
\STATE Set $s=s+1$.
\ENDWHILE
\IF {$\left\Vert \hat{\boldsymbol{\eta}}^{i}-\hat{\boldsymbol{\eta}}^{i-1}\right\Vert \leq\epsilon$ or $i=I$}
\STATE $\hat{\boldsymbol{\eta}}=\hat{\boldsymbol{\eta}}^{i}$.
\STATE  \textbf{break}
\ENDIF

\ENDWHILE
\end{algorithmic}
\label{alg2}
\end{algorithm}

\subsection{Conversion to Position and Clock Offset Estimates}\label{sec:section4c}

While it's possible to estimate the location and clock offset directly from the LOS path geometry, more accurate estimations can be achieved by utilizing the geometry of the NLOS paths. Therefore, following the EXIP theorem, we introduce a weighted least squares formulation to improve the accuracy of localization and clock offset estimation. This approach utilizes estimates from all paths, denoted as $\boldsymbol{\mathbf{\eta}}_\mathrm{p}$,
\begin{equation}
\hat{\boldsymbol{\eta}}_\mathrm{p}=\arg\min_{\boldsymbol{\mathbf{\eta}}_\mathrm{p}}[\hat{\boldsymbol{\mathbf{\eta}}}-f(\boldsymbol{\mathbf{\eta}}_\mathrm{p})]^{\mathsf{T}}\mathbf{F}(\hat{\boldsymbol{\eta}})[\hat{\boldsymbol{\mathbf{\eta}}}-f(\boldsymbol{\mathbf{\eta}}_\mathrm{p})],\label{eq:48}
\end{equation}
where $\mathbf{F}(\hat{\boldsymbol{\eta}})$ is the FIM defined in (\ref{eq:15})
and the mapping $\boldsymbol{\mathbf{\eta}}=f(\boldsymbol{\mathbf{\eta}}_\mathrm{p})$ is described by
(\ref{eq:4}), (\ref{eq:5}), (\ref{eq:6}), (\ref{eq:7}) and (\ref{eq:8}).
The non-linear least squares problem in (\ref{eq:48}) can be solved
via the Nelder-Mead algorithm. The parameters in (\ref{eq:48}) are
initialized with the values $\hat{\mathbf{p}}_{s}$ and $\hat{\Delta}$,
obtained by the following equations
\begin{equation}
\hat{\mathbf{p}}_{s}=\mathbf{p}_{\mathrm{R}}+\hat{d}_{s}\mathbf{k}(\hat{\phi}_{el,s},\hat{\phi}_{az,s}),\label{eq:49}
\end{equation}
\begin{equation}
\hat{\Delta}=\hat{\tau}_{0}-(\hat{d}_{0}+d_{\mathrm{B}})/c.\label{eq:50}
\end{equation}
\begin{remark}
In order to mitigate the negative impact of inaccuracies in multipath information, we employ the estimated clock offset of the LOS path obtained in \eqref{eq:50} as a benchmark to determine the presence of erroneous estimations in the remaining paths, subsequently excluding their contributions in \eqref{eq:48}. Specifically, for the channel parameter estimate of the $s$-th path, compute $\hat{\Delta}_{s}=\hat{\tau}_{s}-(d_{B}+\hat{d}_{s}+\left\Vert \hat{\mathbf{p}}_{s}-\hat{\mathbf{p}}_{0}\right\Vert )/c$.
If the value of $\hat{\Delta}_{s}$ significantly differs from $\hat{\Delta}$, then the estimation information associated with that path should be discarded.
\end{remark}

\section{Discussions}\label{sec:section5}
In this section, we enhance the proposed algorithm to ensure its excellent performance even with a large-scale RIS panel. The improved algorithm is designed to be effective in scenarios with significant near-field curvature, while the originally proposed algorithm remains more efficient for situations with relatively small near-field curvature. Additionally, we have optimized the phase shifts of the RIS, to minimize the CRB on the estimation error.

\subsection{Estimation Algorithm on Ultra-Large RIS}\label{sec:section5a}
\begin{figure}[htb]
  \centering
  \includegraphics[width=7cm]{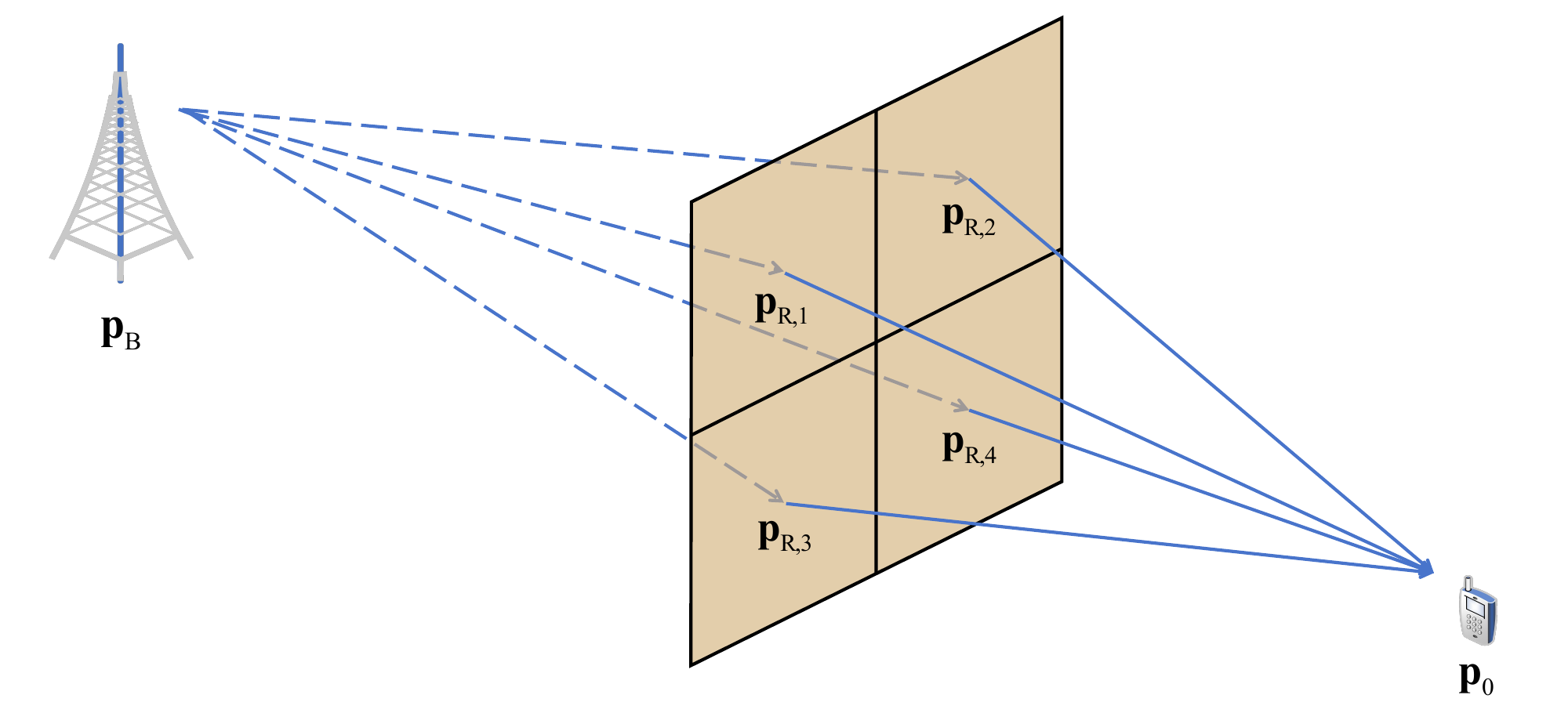}
  \captionsetup{font=small}
  \caption{Division of a UL RIS into $L=4$ sub-RISs}
  \label{fig2}
\end{figure}

Referring to \eqref{eq:2a}, an increase in the aperture size of the RIS can result in an expansion of the near-field range. This, in turn, leads to magnified errors within the far-field approximation of the initial algorithm.
As shown in Fig. \ref{fig2}, to mitigate this effect and reduce the aperture of the RIS panel, the improved algorithm first divides a UL RIS into
$L$
sub-RISs, allowing for the separate estimation of ToA and AoD parameters for each sub-RIS.
\captionsetup{font=small}
\subsubsection{Orthogonal RIS Phase Profiles}

In order to distinguish the transmission signals from each sub-RIS, we need to focus on the design of orthogonal RIS phase profiles, which have been commonly used to differentiate between the LOS path and the path with reflections from the RIS \cite{4JSTSP2022,TWC2021,TVT2021,arXiv2023,TSP2023}.

To start, we divide the overall transmission $T$ into $H\ge L$ blocks,
each containing $\tilde{T}=T/H$ OFDM symbols. We define a matrix $\mathbf{G}\in\mathbb{C}^{H\times L}$
that satisfies the following conditions:
\begin{align}
\mathbf{G}^{\mathsf{T}}\mathbf{G} & =\mathbf{I}_{L\times L},\\
\left|[\mathbf{G}]_{i,j}\right| & =1.
\end{align}

Now, we design the profile matrix $\mathbf{W}_{l,h}\in\mathbb{C}^{\frac{N_{R}}{L}\times\tilde{T}}$
for the $l$-th sub-RIS in the $h$-th block
as follows:
\begin{equation}
\mathbf{W}_{l,h}=[\mathbf{G}]_{h,l}\mathbf{W}_{l},\label{eq:5_A_1_3}
\end{equation}
where $\mathbf{W}_{l}\in\mathbb{C}^{\frac{N_{R}}{L}\times\tilde{T}}$.
Through orthogonal phase profile, the received signal from the $l$-th
sub-RIS can be expressed as:
\begin{equation}
\tilde{\mathbf{Y}}_{\ell}=\frac{1}{H}\sum_{h=1}^{H}[\mathbf{G}]_{h,l}\mathbf{Y}^{h},\label{eq:5_A_1_4}
\end{equation}
where $\mathbf{Y}^{h}\in\mathbb{C}^{N\times\tilde{T}}$ represent
the received signal for the $h$-th block.
\subsubsection{Algorithm Process}
To apply the CPD-OMP algorithm, we design the RIS profile matrix $\mathbf{W}_{l}$ to follow the structure:
\begin{equation}
\mathbf{W}_{l}=\mathbf{T}_{l,1}\otimes\mathbf{T}_{l,2}\in\mathbb{C}^{\frac{N_{x}N_{z}}{L}\times\tilde{T}},\label{eq:56}
\end{equation}
for $l\in{\{1,...,L\}}$, where $\mathbf{T}_{l,1}\in\mathbb{C}^{\frac{N_{x}}{L_{1}}\times\tilde{T}_{1}}$,
$\mathbf{T}_{l,2}\in\mathbb{C}^{\frac{N_{z}}{L_{2}}\times\tilde{T}_{2}}$,
and $L=L_{1}L_{2}$, $\tilde{T}=\tilde{T}_{1}\tilde{T}_{2}$.

Therefore, for the TOA and AOD parameters of the $l$-th sub-RIS, they can be obtained through the CPD-OMP algorithm. Further, coarse estimates of the 3D positions of UE and scatterers, i.e., $\mathbf{p}_{s}$,
can be obtained through least squares as follows \cite{WCL2022}
\begin{equation}
{\hat{\mathbf{p}}_{s}^{c}=\left(\sum_{l=1}^{L} \mathbf{E}_{l,s}\right)^{-1}\left(\sum_{l=1}^{L} \mathbf{E}_{l,s} \mathbf{p}_{\mathrm{R}, l}\right)},\label{eq:5_A_2_2}
\end{equation}
where $\mathbf{E}_{l,s}=\mathbf{I}_{3}-\mathbf{k}(\hat{\phi}_{el,l,s},\hat{\phi}_{az,l,s})\mathbf{k}^{\mathsf{T}}(\hat{\phi}_{el,l,s},\hat{\phi}_{az,l,s})$, and
$\mathbf{p}_{\mathrm{R}, l}$ represents the center of the $l$-th sub-RIS. In the case of a relatively small number of OFDM subcarriers, the time delay resolution is limited, and hence, we obtain a coarse estimate of clock offset using the following expression
\begin{equation}
\hat{\Delta}^{c}=\frac{1}{L}\sum_{l=1}^{L}\hat{\tau}_{l}-(\left\Vert \hat{\mathbf{p}}_{0}-\mathbf{p}_{\mathrm{R}}\right\Vert +d_{\mathrm{B}})/c.\label{eq:5_A_2_3}
\end{equation}

The overall estimation algorithm on UL RIS  is shown in Algorithm 3.
\begin{algorithm}[H]
\caption{Estimation algorithm on UL RIS}\label{alg:alg3}
\renewcommand{\algorithmicrequire}{\textbf{Input:}}
\renewcommand{\algorithmicensure}{\textbf{Output:}}
\begin{algorithmic}[1]
\REQUIRE Recieved signals matrix $\mathbf{Y}$, RIS profile matrix $\mathbf{W}_{l}$ and the matrix $\mathbf{G}$ in \eqref{eq:5_A_1_3}.
\ENSURE $\hat{\mathbf{p}}_{0}$, $\hat{\Delta}$.
\STATE The received signal $\tilde{\mathbf{Y}}_{\ell}$
from each sub-RIS path can be separated by using \eqref{eq:5_A_1_4}.
\STATE For the $l$-th sub-RIS, apply Algorithm 1 to estimate
$\{{\tau}_{l,s}\}_{s=0}^{N_{s}}$, $\{{\phi}_{el,l,s}\}_{s=0}^{N_{s}}$, and $\{{\phi}_{az,l,s}\}_{s=0}^{N_{s}}$.
\STATE Obtain $\hat{\mathbf{p}}_{s}^{c}$ using \eqref{eq:5_A_2_2},
then obtain $\hat{\Delta}^{c}$ using \eqref{eq:5_A_2_3}.
\STATE Obtain coarse estimate of $\boldsymbol{\eta}$ by \eqref{eq:4}, \eqref{eq:5}, \eqref{eq:6}, \eqref{eq:7},
\eqref{eq:8}, and \eqref{eq:42}.
\STATE Refine channel parameters using Algorithm 2.
\STATE Conversion to $\hat{\mathbf{p}}_{0}$ and $\hat{\Delta}$ according to EXIP.
\end{algorithmic}
\label{alg3}
\end{algorithm}

\subsection{Optimization of RIS Phase Shifts}\label{sec:section5b}

In this section, we aim to optimize the phase profile of the RIS by minimizing the PEB.
Achieving an optimal phase profile for the RIS usually demands prior knowledge about the target's location. To address this, we suggest leveraging the position information acquired through the estimation algorithm we introduced as prior information for phase optimization. This strategic approach proves effective in reducing PEB. Consequently, applying Algorithm 3 on its own subsequently leads to a significant improvement in localization accuracy.

\subsubsection{Problem Formulation}

We begin by rewritten the FIM in (\ref{eq:15}):
\begin{equation}
\mathbf{F}(\boldsymbol{\eta})=\frac{2}{\sigma^{2}}\sum_{n=1}^{N}\mathfrak{R}\left\{ \left(\frac{\partial\boldsymbol{\mu}[n]}{\partial\boldsymbol{\eta}}\right)^{\mathrm{\mathsf{H}}}\left(\frac{\partial\boldsymbol{\mu}[n]}{\partial\boldsymbol{\eta}}\right)\right\}.\label{eq:59}
\end{equation}

Here,
\begin{align}
\boldsymbol{\mu}[n]&=\mathbf{W}^{\mathsf{T}}\sqrt{P}\sum_{s=0}^{N_{s}}\rho_{s}e^{-j2\pi\tau_{s}(n-1)\varDelta f}\mathbf{b}\left(\mathbf{p}_{s}\right)\nonumber \\
&\triangleq\mathbf{W}^{\mathsf{T}}\boldsymbol{\kappa}[n],
\end{align}
and $\boldsymbol{\kappa}[n]\triangleq\sqrt{P}\sum_{s=0}^{N_{s}}\rho_{s}e^{-j2\pi\tau_{s}(n-1)\varDelta f}\mathbf{b}\left(\mathbf{p}_{s}\right)$.

We can further represent $\partial\boldsymbol{\mu}[n]/\partial\boldsymbol{\eta}_{s}$
as follows:
\begin{align}
\frac{\partial\boldsymbol{\mu}[n]}{\partial\boldsymbol{\eta}_{s}} &= \mathbf{W}^{\mathsf{T}}
\left[\frac{\partial\boldsymbol{\kappa}[n]}{\partial\mathfrak{R}(\rho_{s})},\frac{\partial\boldsymbol{\kappa}[n]}{\partial\mathfrak{I}(\rho_{s})},\right. \nonumber \\
&\left.\frac{\partial\boldsymbol{\kappa}[n]}{\partial\phi_{el,s}},\frac{\partial\boldsymbol{\kappa}[n]}{\partial\phi_{az,s}},\frac{\partial\boldsymbol{\kappa}[n]}{\partial d_{s}},\frac{\partial\boldsymbol{\kappa}[n]}{\partial\tau_{s}}\right] \nonumber \\
&\triangleq \mathbf{W}^{\mathsf{T}}\mathbf{K}_{s}[n],\label{eq:60}
\end{align}
their derivations are given in Appendix C.
Through defining
\begin{equation}
\mathbf{K}[n]=[\mathbf{K}_{0}[n],...,\mathbf{K}_{N_{s}}[n]],
\end{equation}
we can represent $\partial\boldsymbol{\mu}[n]/\partial\boldsymbol{\eta}$
as
\begin{equation}
\frac{\partial\boldsymbol{\mu}[n]}{\partial\boldsymbol{\eta}}=\mathbf{W}^{\mathsf{T}}\mathbf{K}[n].
\end{equation}

Now, we can rewrite (\ref{eq:59}) as
\begin{equation}
\begin{aligned}
\mathbf{F}(\boldsymbol{\eta}) &= \frac{2}{\sigma^2} \sum_{n=1}^{N} \mathfrak{R}\left\{ \mathbf{K}^{\mathrm{\mathsf{H}}}[n] (\mathbf{W}\mathbf{W}^{\mathsf{H}})^* \mathbf{K}[n] \right\} \\
&\triangleq \frac{2}{\sigma^2} \sum_{n=1}^{N} \mathfrak{R}\left\{ \mathbf{K}^{\mathrm{\mathsf{H}}}[n] \mathbf{\Lambda}^* \mathbf{K}[n] \right\},
\end{aligned}
\end{equation}
where $\mathbf{\Lambda}\triangleq\mathbf{W}\mathbf{W}^{\mathsf{H}}$.
According to (\ref{eq:17}), we obtain
\begin{equation}
\mathbf{F}(\boldsymbol{\eta}_{p})=\frac{2}{\sigma^{2}}\sum_{n=1}^{N}\mathbf{J}\mathfrak{R}\left\{ \mathbf{K}^{\mathrm{\mathsf{H}}}[n]\mathbf{\Lambda}^{*}\mathbf{K}[n]\right\} \mathbf{J}^{\mathsf{T}}.\label{eq:5B7}
\end{equation}
\begin{remark}
In practical operations, we replace $\mathbf{F}(\boldsymbol{\eta}_\mathrm{p})$ with
$\mathbf{F}(\boldsymbol{\bar{\eta}}_\mathrm{p})$ in \eqref{eq:5B7}, where
$\boldsymbol{\bar{\eta}}_\mathrm{p}\triangleq[\mathbf{p}^{\mathsf{T}},\Delta]^{\mathsf{T}}\in\mathbb{R}^{3N_{s}+4}$.
This substitution significantly reduces the dimension of the FIM and substantially improves the numerical stability.
\end{remark}

Based on the definition of PEB in Section \ref{sec:section3b}, the RIS phase
optimization problem can be formulated as
\begin{subequations}\label{eq:65}
\begin{align}
\underset{\mathbf{W}}{\min} \quad & \mathrm{tr}[\mathbf{F}^{-1}(\boldsymbol{\eta}_\mathrm{p})]_{1:3,1:3} \label{eq:65a}\\
\text{s.t.} \quad & \mathbf{\Lambda} \succeq 0, \label{eq:65b}\\
& \mathrm{diag}(\mathbf{\Lambda}) = T, \label{eq:65c}\\
& \mathrm{rank}(\mathbf{\Lambda}) \leq T.\label{eq:65d}
\end{align}
\end{subequations}

By applying Schur complement, the optimization problem \eqref{eq:65} can be
reformulated as
\begin{subequations}
\begin{align}
\underset{\mathbf{t},\mathbf{W}}{\min} \quad & \mathbf{1}^{\mathsf{T}}\mathbf{t} \\
\text{s.t.} \quad & \left[\begin{array}{cc}
\mathbf{F}(\boldsymbol{\eta}_\mathrm{p}) & \mathbf{e}_{k} \\
\mathbf{e}_{k}^{\mathsf{T}} & t_{k}
\end{array}\right] \succeq 0, k=1,2,3, \label{eq:66b}\\
& \eqref{eq:65b},\eqref{eq:65c},\eqref{eq:65d}, \nonumber
\end{align}
\end{subequations}
where $\mathbf{t}=[t_{1},t_{2},t_{3}]^{\mathsf{T}}$ is an auxiliary
variable and $\mathbf{e}_{k}$ is the $k$-th column of the identity
matrix. From \eqref{eq:5B7}, it's evident that $\mathbf{F}(\boldsymbol{\eta}_{p})$
is a linear function of $\mathbf{\Lambda}$, but quadratic in $\mathbf{W}$.
Consequently, we switch the optimization variable to $\mathbf{\Lambda}$,
converting constraint \eqref{eq:65b} into a linear matrix inequality (LMI) constraint.
At the same time, we drop the non-convex constraint \eqref{eq:65d} since it is always satisfied.
This leads to the reformulated optimization problem:
\begin{align}
\underset{\mathbf{t},\mathbf{\Lambda}}{\min} \quad & \mathbf{1}^{\mathsf{T}}\mathbf{t} \\
\text{s.t.} \quad &\eqref{eq:66b},\eqref{eq:65b},\eqref{eq:65c}. \nonumber
\end{align}

This optimization problem is a convex semidefinite program (SDP) \cite{boyd2004}.
However, it's worth noting that the optimization variable
$\mathbf{\Lambda}\in\mathbb{C}^{N_\mathrm{R}\times{N_\mathrm{R}}}$ has a high dimension.
In the next subsection, we will introduce a method with lower complexity to solve this optimization problem.

\subsubsection{Solve Problem of RIS Phase Design}\label{sec:section5b2}

By means of \eqref{eq:98}, we can obtain matrix $\mathbf{B}_{s}$,
which serves as the basis for the column space of $\mathbf{K}_{s}[n]$,
and it is defined as follows:
\begin{equation}
\mathbf{B}_{s}\triangleq[\mathbf{b}(\mathbf{p}_{s}),\dot{\mathbf{b}}_{\phi_{el,s}}(\mathbf{p}_{s}),\dot{\mathbf{b}}_{\phi_{el,s}}(\mathbf{p}_{s}),\dot{\mathbf{b}}_{d_{s}}(\mathbf{p}_{s})]\in\mathbb{C}^{N_{\mathrm{R}}\times4},\label{eq:5_B_2_1}
\end{equation}
where $\dot{\mathbf{b}}_{x}(\mathbf{p}_{s})\triangleq\partial\mathbf{b}(\mathbf{p}_{s})/\partial x$.
Furthermore, we define
\begin{equation}
\mathbf{B}\triangleq[\mathbf{B}_{0},...,\mathbf{B}_{N_{s}}]\in\mathbb{C}^{N_{\mathrm{R}}\times4(N_{s}+1)}.\label{eq:5_B_2_2}
\end{equation}

It is evident that $\mathbf{K}[n]\in\mathcal{S}(\mathbf{B})$, for $n\in\{1,...,N\}$,
where $\mathcal{S}(\mathbf{B})$ denotes the column space of $\mathbf{B}$.

\begin{proposition}[\cite{TSP2008,7JSTSP2022}]\label{prop:prop1}
The optimal RIS phase profile covariance matrix $\mathbf{\Lambda}$
in the absence of the unit modulus constraints $\mathrm{diag}(\mathbf{\Lambda}) = T$
can be expressed as
\begin{equation}
\mathbf{\Lambda}=(\mathbf{B}\mathbf{\Xi}\mathbf{B}^{\mathsf{H}})^{*},\label{eq:5_B_2_3}
\end{equation}
where $\mathbf{\Xi}\in\mathbb{C}^{4(N_{s}+1)\times4(N_{s}+1)}$ is
a positive semidefinite matrix.
\end{proposition}

By applying Proposition \ref{prop:prop1}, the complexity of the optimization problem is significantly reduced.
However, it's important to note that
Proposition \ref{prop:prop1} does not account for constraint \eqref{eq:65c},
we slightly relax constraint \eqref{eq:65c} and reformulate the optimization problem as follows
\begin{align}
\underset{\mathbf{t},\mathbf{\Xi}}{\mathrm{min}} \quad & \mathbf{1}^{\mathsf{T}}\mathbf{t}+\gamma\left\|\mathrm{diag}(\mathbf{\Lambda})-T\right\| \label{eq:5_B_2_4}\\
\text{s.t.} \quad
& \eqref{eq:66b},\eqref{eq:65b}, \nonumber
\end{align}
where $\gamma$ is a regularization parameter.
We can solve the problem in \eqref{eq:5_B_2_4} using existing convex solvers \cite{CVX}.
Once $\mathbf{\Xi}$ is obtained, it can be further used to calculate $\mathbf{\Lambda}$ as \eqref{eq:5_B_2_3}.
Now, to generate RIS phase profiles that satisfy the unit modulus constraint,
we will introduce a Gaussian randomization-based method \cite{18GLOBECOM2018,TSP2006} to derive the RIS profile matrix $\mathbf{W}$.
This method is outlined in Algorithm 4.
\begin{algorithm}[H]
\caption{Gaussian randomization method to derive $\mathbf{W}_{opt}$}\label{alg:alg4}
\renewcommand{\algorithmicrequire}{\textbf{Input:}}
\renewcommand{\algorithmicensure}{\textbf{Output:}}
\begin{algorithmic}[1]
\REQUIRE $\mathbf{\Xi}_{opt}$ from \eqref{eq:5_B_2_4}, $\mathbf{B}$ from \eqref{eq:5_B_2_2}.
\ENSURE RIS profile matrix $\mathbf{W}_{opt}$.
\STATE Calculate $\mathbf{\Lambda}_{opt}$ by using \eqref{eq:5_B_2_3}.
\STATE Perform eigenvalue decomposition of $\mathbf{\Lambda}_{opt}$:
$\mathbf{\Lambda}_{opt}=\mathbf{U\Sigma U}^{\mathsf{H}}$.
\STATE Generate $\mathbf{R}\in\mathbb{C}^{N_{R}\times T}$, with its entries are distributed as
$\mathcal{CN}(0,1)$. Obtain $\mathbf{\widetilde{W}=}\mathbf{U\Sigma}^{1/2}\mathbf{R}$.
\STATE Compute $\mathbf{W}_{opt}=\mathrm{exp}[j\mathrm{arg}([\frac{\widetilde{\mathbf{W}}}{[\widetilde{\mathbf{W}}]_{N_{R},T}}])]$.
\end{algorithmic}
\label{alg4}
\end{algorithm}

\subsubsection{Proposed RIS Phase Design}\label{sec:section5b3}
The method proposed in Section \ref{sec:section5b2} significantly reduces the number of optimization variables. However, the number of optimization variables still increases linearly with the number of paths, leading to higher complexity. 
To further address this complexity, we approximate the matrix $\mathbf{\Xi}$ in the \eqref{eq:5_B_2_3} as a block diagonal matrix, leading to the approximation:
\begin{equation}
\mathbf{\Lambda}\thickapprox\sum_{s=0}^{N_{s}}\lambda_{s}(\mathbf{B}_{s}\mathbf{\Xi}_{s}\mathbf{B}_{s}^{\mathsf{H}})^{*}\triangleq\sum_{s=0}^{N_{s}}\lambda_{s}\mathbf{\Lambda}_{s},\label{eq:5_B_3_1}
\end{equation}
where $\mathbf{\Xi}_{s}\in\mathbb{C}^{4\times4}$. Thus, the optimization problem \eqref{eq:5_B_2_4} can be approximated as $N_{s}+1$ sub-problems as follows
\begin{subequations}
\begin{align}
\underset{\mathbf{t},\mathbf{\Xi}_{s}}{\mathrm{min}} \quad & \mathbf{1}^{\mathsf{T}}\mathbf{t}+\gamma\left\Vert \mathrm{diag}(\mathbf{\Lambda}_{s})-T\right\Vert \\
\text{s.t.} \quad & \left[\begin{array}{cc}
\mathbf{F}(\boldsymbol{\eta}_{\mathrm{p}_{s}}) & \mathbf{e}_{k}\\
\mathbf{e}_{k}^{\mathsf{T}} & t_{k}
\end{array}\right]\succeq0,k=1,2,3,\label{eq:5_B_3_2}\\
& \eqref{eq:65b},\nonumber
\end{align}
\end{subequations}
where $\mathbf{F}(\boldsymbol{\eta}_{\mathrm{p}_{s}})=\frac{2}{\sigma^{2}}\sum_{n=1}^{N}\mathbf{J}_{s}\mathfrak{R}\left\{ \mathbf{K}_{s}^{\mathrm{\mathsf{H}}}[n]\mathbf{\Lambda}_{s}^{*}\mathbf{K}_{s}[n]\right\} \mathbf{J}_{s}^{\mathsf{T}}$, $\mathbf{J}_{s}=\partial\boldsymbol{\eta}_{s}^{\mathsf{T}}/\partial\boldsymbol{\eta}_{\mathrm{p}_{s}}$ and $\boldsymbol{\eta}_{\mathrm{p}_{s}}\triangleq[\mathbf{p}_{s}^{\mathsf{T}},\Delta]^{\mathsf{T}}$. 
Each sub-optimization problem can be viewed as minimizing the PEB at the given position $\mathbf{p}_{s}$, which can be efficiently solved using existing convex solvers. Subsequently, by solving the subsequent optimization problem, we can determine the values of the weight vector  
$\boldsymbol{\lambda}$
\begin{align}
\underset{\mathbf{t},\boldsymbol{\lambda}}{\mathrm{min}} \quad& \mathbf{1}^{\mathsf{T}}\mathbf{t}+\gamma\left\Vert \mathrm{diag}(\mathbf{\Lambda})-T\right\Vert \\
\text{s.t.} \quad & \eqref{eq:66b},\eqref{eq:65b},\nonumber
\end{align}
where $\boldsymbol{\lambda}=[\lambda_{0},...,\lambda_{N_{s}}]^{\mathsf{T}}$. Finally, we can calculated $\boldsymbol{\Lambda}$ using \eqref{eq:5_B_3_1}, and derive the RIS phase profile matrix $\mathbf{W}$ by applying Algorithm 4.
\section{Numerical Results}\label{sec:section6}
In this section, we present numerical results to evaluate the
theoretical bounds derived in Section \ref{sec:section3} and the performance of
the proposed algorithms in Section \ref{sec:section4} and Section \ref{sec:section5}.

\begin{figure}
\centering

\begin{minipage}{0.5\textwidth}
  \centering

  \includegraphics[width=1\linewidth]{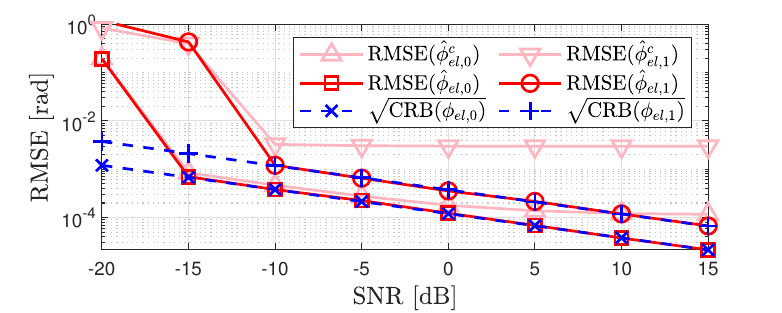}
  \centerline{\fontsize{9}{11}\selectfont(a) The RMSE of elevation angle estimation.}
\end{minipage}

\begin{minipage}{0.5\textwidth}
  \centering
  \includegraphics[width=1\linewidth]{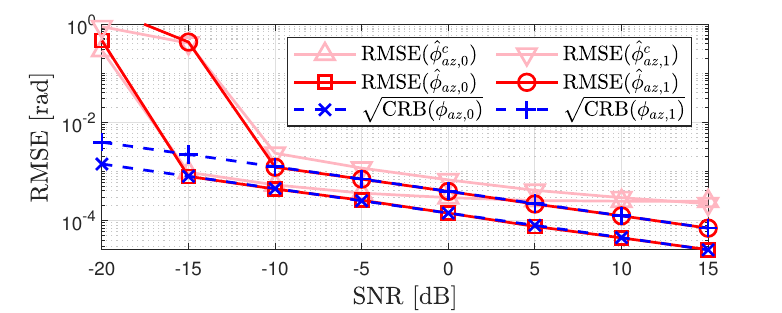}
  \centerline{\fontsize{9}{11}\selectfont(b) The RMSE of azimuth angle estimation.}

\end{minipage}

\begin{minipage}{0.5\textwidth}
  \centering
  \includegraphics[width=1\linewidth]{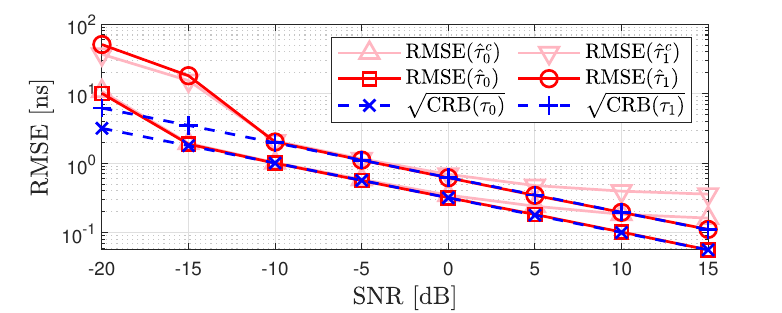}
  \centerline{\fontsize{9}{11}\selectfont(c) The RMSE of TOA estimation.}

\end{minipage}
\begin{minipage}{0.5\textwidth}
  \centering
  \includegraphics[width=1\linewidth]{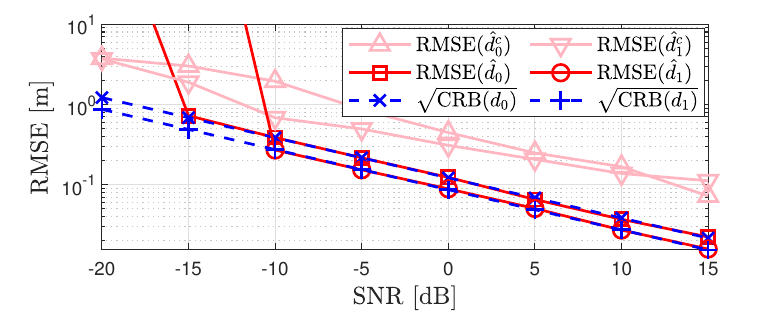}
  \centerline{\fontsize{9}{11}\selectfont(d) The RMSE of distance estimation.}

\end{minipage}
\captionsetup{justification=raggedright,singlelinecheck=false}
\captionsetup{font=small}
\caption{The RMSE of channel parameters versus SNR.}
\label{fig3}
\end{figure}

\begin{figure}
\centering

\begin{minipage}{0.5\textwidth}
  \centering
  \includegraphics[width=1\linewidth]{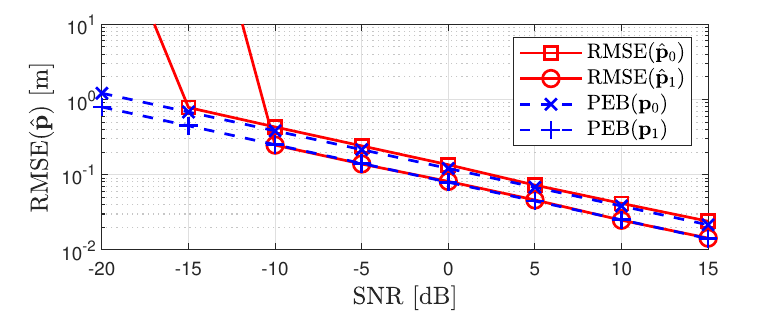}
  \centerline{\fontsize{9}{11}\selectfont(a) The RMSE of position estimation.}
\end{minipage}

\begin{minipage}{0.5\textwidth}
  \centering
  \includegraphics[width=1\linewidth]{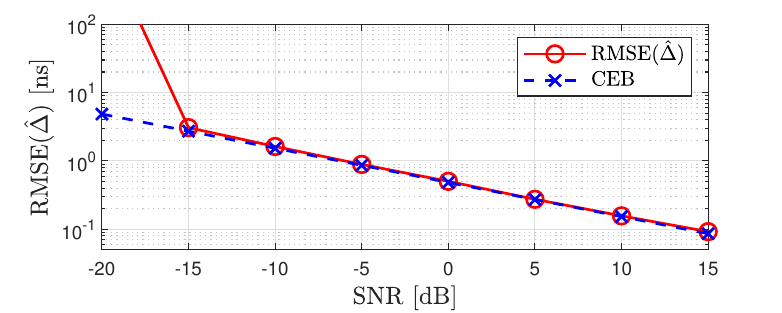}
  \centerline{\fontsize{9}{11}\selectfont(b) The RMSE of clock offset estimation.}

\end{minipage}
\captionsetup{font=small}
\caption{The RMSE of position and clock offset estimations using Proposed 1 versus SNR.}
\label{fig4}
\end{figure}
\begin{figure}
\centering

\begin{minipage}{0.5\textwidth}
  \centering
  \includegraphics[width=1\linewidth]{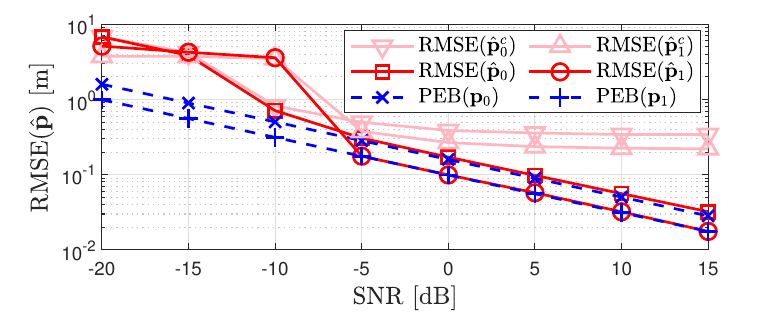}
  \centerline{\fontsize{9}{11}\selectfont(a) The RMSE of position estimations.}
\end{minipage}

\begin{minipage}{0.5\textwidth}
  \centering
  \includegraphics[width=1\linewidth]{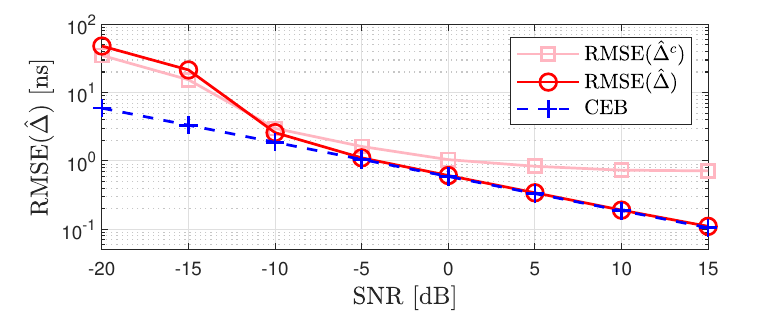}
  \centerline{\fontsize{9}{11}\selectfont(b) The RMSE of clock offset estimation.}

\end{minipage}
\captionsetup{font=small}
\caption{The RMSE of position and clock offset estimations using Proposed 2 versus SNR.}
\label{fig5}
\end{figure}
\subsection{Simulation Setup}
We consider an indoor localization scenario within a 10m\texttimes 10m\texttimes 4m
space. The channel gains of the BS-RIS, RIS-UE and RIS-scatterer-UE
paths are generated with $\rho_{\mathrm{BR}}=\lambda/(4\pi d_{\mathrm{B}})e^{j\alpha_{\mathrm{B}}}$,
$\rho_{\mathrm{RU},0}=\lambda/(4\pi d_{0})e^{j\alpha_{0}}$, $\rho_{\mathrm{RU},s}=\kappa\lambda/(4\pi(d_{s}+\left\Vert \mathbf{p}_{0}-\mathbf{p}_{s}\right\Vert )e^{j\alpha_{s}}$,
where $\kappa$ is the reflection loss and $\alpha_{\mathrm{B}}$,
$\alpha_{0}$ and $\alpha_{s}$ are independently generated from a
uniform distribution $\mathcal{U}(0,2\pi)$.
Unless specified otherwise, we set $f_{c}=28\mathrm{GHz}$,
$\varDelta f=120\mathrm{kHz}$, $c=3\times10^{8}\mathrm{m/s}$, $P=29\mathrm{dBm}$, $\sigma^{2}=-115.2\mathrm{dBm}$, $N=80$, $T=256$,
$\kappa=0.6$, $N_{x}=N_{z}=48$ and $\Delta=100\mathrm{ns}$.
The BS is located at $\mathbf{p}_{\mathrm{B}}=[0,-60,5]^{\mathsf{T}}$
m, the RIS is located at $\mathbf{p}_{\mathrm{R}}=[0,0,0]^{\mathsf{T}}$
m, and the UE is located at $\mathbf{p}_{0}=[3,6,-1]^{\mathsf{T}}$ m.
Assume that there is a scatterer in the link from the RIS to the UE
with location $\mathbf{p}_{1}=[-1,3,2]^{\mathsf{T}}$ m.
In addition, the SNR is defined as
\begin{equation}
\mathrm{SNR}=\frac{\sum_{t=1}^{T}\sum_{n=1}^{N}\left|\mu_{t}[n]\right|^{2}}{\sigma^{2}NT}.
\end{equation}

In all simulation examples, the
RMSEs are computed over 1000 Monte Carlo trials.
The proposed algorithm in Section \ref{sec:section4} is labeled as “Proposed 1” and the proposed algorithm in Section \ref{sec:section5a} is labeled as “Proposed 2”. Additionally, the RIS phase profile designed in Section \ref{sec:section5b2} is labeled as “Optimized 1” and the RIS phase profile designed in Section \ref{sec:section5b3} is labeled as “Optimized 2”.

\subsection{Accuracy of Channel Parameters}
The RMSEs of AOD, TOA and distance estimation are shown in Fig. \ref{fig3}(a)-(d), respectively.
Here, $\hat{\phi}_{el,s}^{c}$, $\hat{\phi}_{el,s}^{c}$, $\hat{\tau}_{s}^{c}$, and $\hat{d}_{s}^{c}$ are coarsely-estimated
by Algorithm 1, while $\hat{\phi}_{el,s}$, $\hat{\phi}_{el,s}$, $\hat{\tau}_{s}$, and $\hat{d}_{s}$ are refined by Algorithm 2.
When the scatterer is positioned closer to the RIS, it results in a larger near-field curvature, thereby leading to more precise distance estimation. However, due to the increased path loss along the scatterer's path, the accuracy in estimating other channel parameters is not as high as that of the UE path.
Notably, when the SNR exceeds -15 dB, channel parameter estimation for the UE approaches the performance boundary. Furthermore, when the SNR surpasses -10 dB, the estimation of channel parameters for the scatterer also demonstrates its ability to reach the performance boundary.
By comparing coarse and refined estimations, it becomes evident that under extremely low SNR conditions, inaccurate coarse estimates can lead to divergence in the refined results. Moreover, the coarse estimation accuracy, relying on the far-field approximation, saturates relatively quickly with SNR improvement, consistent with conclusions in existing literature\cite{GLOBECOM2022}. Fortunately, simulations suggest that the refinement process can rectify errors introduced by the far-field approximation.

\begin{figure}[t]
\centering
\begin{minipage}{0.5\textwidth}
  \centering
  \includegraphics[width=1\linewidth]{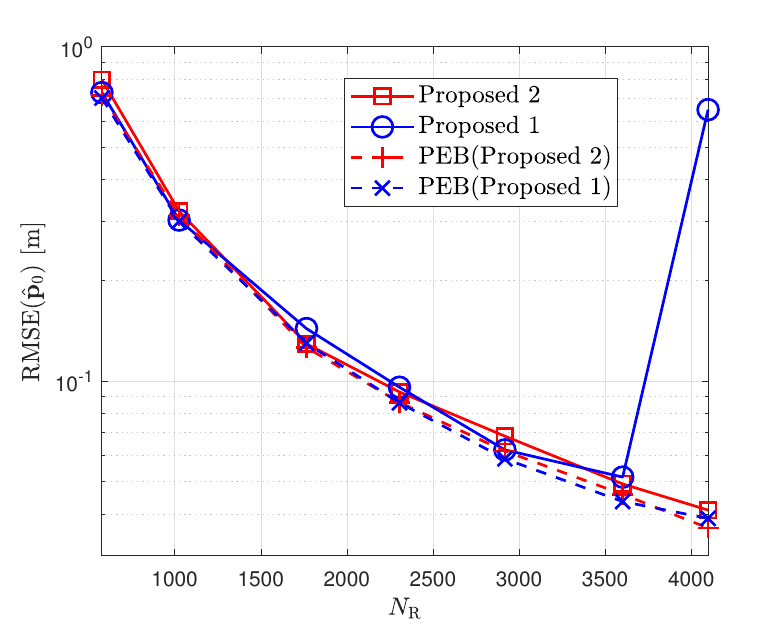}

\end{minipage}
\captionsetup{justification=raggedright,singlelinecheck=false}
\caption{The RMSE of UE position versus number of RIS elements.}
\label{fig6}
\end{figure}

\subsection{Accuracy of Positioning and Clock Offset}

In this section, we apply the methodology detailed in Section \ref{sec:section4c} to translate the channel parameter estimates into predictions for both location and clock offset. The RMSE results for target (UE and scatterer) positions and clock offset using Proposed 1 are presented in Fig. \ref{fig4}.
Thanks to the more accurate distance estimation, the position error of the scatterer is smaller than that of the UE. It's worth noting that, at an SNR of -15dB, we exclusively rely on channel parameters from the UE path to determine the clock offset  and UE's position. This is due to the significant inaccuracy in estimating channel parameters related to the scatterer under these conditions. In this scenario, even though multipath information cannot be distinguished, relatively accurate estimates of the UE's position and clock offset can still be obtained.
Furthermore, it's evident that the estimates for locations and clock offset approach the theoretical performance boundary.

Fig. \ref{fig5} depicts the RMSE of positions and clock offset estimates obtained through Proposed 2.
Here, $\hat{\mathbf{p}}_{s}^{c}$ is coarsely-estimated
according to \eqref{eq:5_A_2_2},
and $\hat{\Delta}^{c}$ by \eqref{eq:5_A_2_3}.
Comparing Fig. \ref{fig4} and Fig. \ref{fig5}, it can be observed that at low SNRs, Proposed 1 exhibits superior performance, while at higher SNRs, both Proposed 1 and Proposed 2 nearly achieve the performance bounds.
Overall, both methods' coarse estimates can provide good initial values, enabling the refined algorithm based on SAGE to converge effectively.
\begin{figure}[t]
  \centering
  \includegraphics[width=1\linewidth]{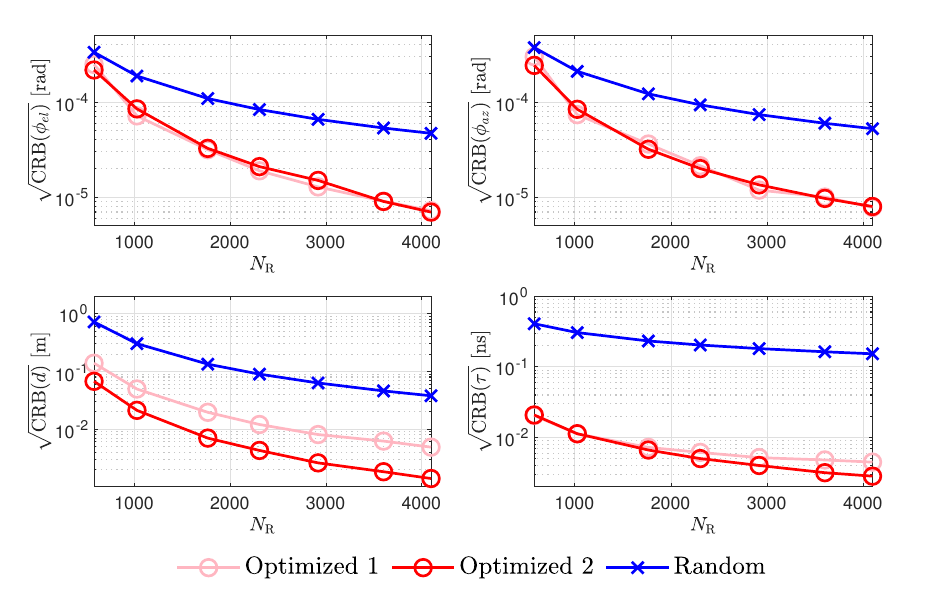}
  \caption{CRB of channel parameters versus number of RIS elements for different RIS phase design strategies.}
  \label{fig8}
\end{figure}
\begin{figure}[t]
  \centering
  \includegraphics[width=1\linewidth]{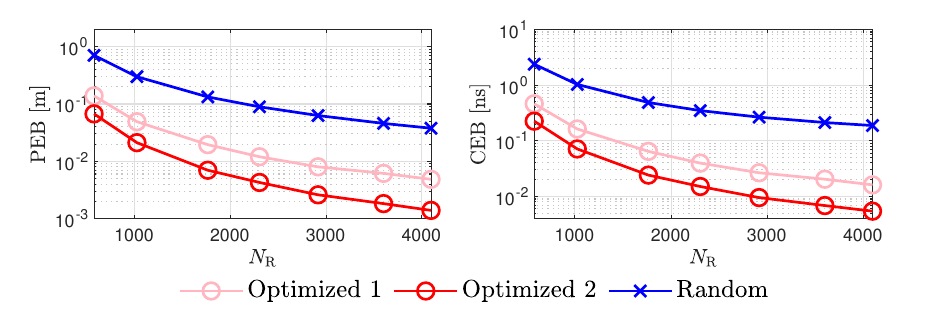}
  \caption{PEB and CEB versus number of RIS elements for different RIS phase design strategies.}
  \label{fig7}
\end{figure}
\begin{figure}[ht]
  \centering
  \includegraphics[width=1\linewidth]{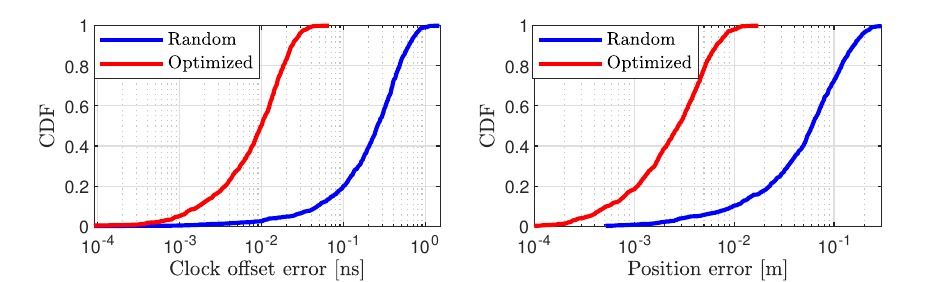}
  \caption{CDF of position error and clock offset error for different RIS phase design strategies.}
  \label{figcdf}
\end{figure}

\subsection{Comparison between Proposed 1 and Proposed 2}

Fig. \ref{fig6} provides a comparison of the estimation performance between Proposed 1 and Proposed 2 across various sizes of RIS. As Proposed 1 and Proposed 2 have distinct requirements for RIS phase profiles, their respective performance boundaries also differ.
Evidently, when $N_{\mathrm{R}}=4096$, the performance of Proposed 1 notably deteriorates. This indicates that under extreme near-field effects conditions, the far-field approximation is no longer capable of providing an effective initial angle value for the near-field model.
Meanwhile, the performance of Proposed 2 can still approach the performance limit, demonstrating the effectiveness of Proposed 2 in scenarios with extremely large RIS-assisted positioning.
In most other cases, there is no significant disparity in the estimation performance between them, and their estimation accuracy can approach the performance boundary.
Based on this simulation result, we recommend using Proposed 2 in scenarios with extremely large RIS deployment, and utilizing Proposed 1 for estimation in other cases. The reason for not prioritizing Proposed 2 is that it has stricter requirements for the number of OFDM symbols, and its performance is not as strong as Proposed 1 under low SNR conditions.
\begin{figure}[t]
\centering
\begin{minipage}[b]{1\linewidth}
  \centering
  \includegraphics[width=1\linewidth]{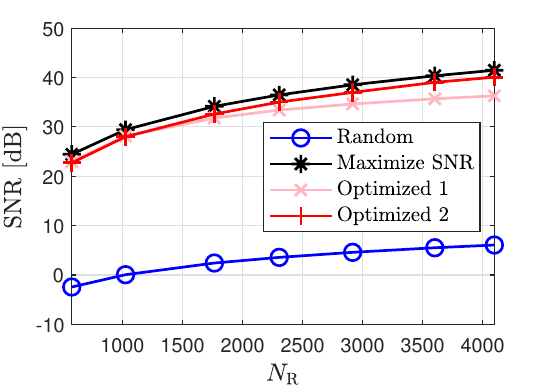}
\end{minipage}
\captionsetup{justification=raggedright,singlelinecheck=false}
\caption{SNR versus number of RIS elements for different RIS phase design strategies.}
\label{fig9}
\end{figure}
\subsection{RIS Phase Shifts Optimization}
Next, we assess the RIS phase shifts optimization performance introduced in Section \ref{sec:section5b}.
In Fig. \ref{fig8}, we illustrate the CRB for various channel parameters under the condition of minimizing the PEB and in Fig. \ref{fig7} we depict PEB and CEB versus number of RIS elements for different RIS phase design strategies. Notably, the optimization method presented in Section \ref{sec:section5b3} demonstrates superior results, particularly evident in the CRB for distance, which further impacts the PEB and CEB. This could be due to numerical instability or improper regularization parameter settings in the original optimization problem \eqref{eq:5_B_2_4}. Conversely, the approximate optimization problem, achieved through distributed computing and exploiting the original problem's special structure, yields better results. Moreover, across different RIS panel sizes, optimized channel parameters consistently exhibit substantial performance enhancements.

Fig. \ref{figcdf} presents the cumulative distribution functions (CDFs)
of the estimation error for 1000 different realizations
of random RIS phase profile and optimized RIS phase profile (presented in Section \ref{sec:section5b3}). 
The parameter estimation process under the optimized phase profile involves using the estimated channel parameters under random RIS phase profile as initial values, followed by applying Algorithm 3. It is evident that the proposed RIS design method has significantly improved the accuracy of localization and synchronization by more than an order of magnitude compared to random RIS phase designs. Furthermore, in Fig. \ref{fig9}, the change in SNR for different RIS phase design stratrgies is displayed. It can be observed that the proposed RIS design has greatly improved the SNR, and it is close to the maximum SNR. This indicates that the phase configuration, while enhancing positioning accuracy, also enhances communication performance.

%

%
%
%

\section{Conclusion}\label{sec:section7}
In this paper, the problem of RIS-aided 3D localization and synchronization in multipath environments has been studied, focusing on the near-field of mmWave systems.
We have introduced two novel positioning frameworks, tailored to scenarios with varying degrees of near-field effects. The first framework leverages tensor representation, CPD, and principles of compressed sensing to estimate channel parameters. Subsequently, the estimation is refined using the SAGE algorithm, with the final step involving the conversion of channel parameter estimates into position and clock offset estimates through weighted least squares.
Building upon this foundation, the second framework takes a step further by transforming a problem involving an UL RIS-assisted positioning into one with multiple simultaneous sub-RISs positioning challenges. This transformation is achieved through the design of orthogonal phase contours, which helps reduce errors associated with far-field approximations.
In addition, to further enhance positioning accuracy, we have also optimized the phase profile of the RIS. Finally, simulation experiments validated the effectiveness of the proposed algorithms.


{\appendices
\section{FIM of the channel parameter}
For convenience, we rewrite \eqref{eq:16} as follows:
\begin{equation}
\mu_{t}[n]=\sqrt{P}\sum_{s=0}^{N_{s}}\rho_{s}e^{-j2\pi\tau_{s}(n-1)\varDelta f}\mathbf{g}_{t}^{\mathsf{T}}\mathbf{a}\left(\mathbf{p}_{s}\right),
\end{equation}
where $\mathbf{g}_{t}\triangleq\mathbf{a}\left(\mathbf{p}_{\mathrm{B}}\right)\odot\mathbf{w}_{t}$.
Then we can obtain the derivatives as follows:
\begin{align}
\frac{\partial\mu_{t}[n]}{\partial\mathfrak{R}(\rho_{s})} & =\sqrt{P}e^{-j2\pi\tau_{s}(n-1)\varDelta f}\mathbf{g}_{t}^{\mathsf{T}}\mathbf{a}\left(\mathbf{p}_{s}\right),\\
\frac{\partial\mu_{t}[n]}{\partial\mathfrak{I}(\rho_{s})} & =j\sqrt{P}e^{-j2\pi\tau_{s}(n-1)\varDelta f}\mathbf{g}_{t}^{\mathsf{T}}\mathbf{a}\left(\mathbf{p}_{s}\right),\\
\frac{\partial\mu_{t}[n]}{\partial\phi_{el,s}} & =\sqrt{P}\rho_{s}e^{-j2\pi\tau_{s}(n-1)\varDelta f}\mathbf{g}_{t}^{\mathsf{T}}\dot{\mathbf{a}}_{\phi_{el,s}}\left(\mathbf{p}_{s}\right),\\
\frac{\partial\mu_{t}[n]}{\partial\phi_{az,s}} & =\sqrt{P}\rho_{s}e^{-j2\pi\tau_{s}(n-1)\varDelta f}\mathbf{g}_{t}^{\mathsf{T}}\dot{\mathbf{a}}_{\phi_{az,s}}\left(\mathbf{p}_{s}\right),\\
\frac{\partial\mu_{t}[n]}{\partial d_{s}} & =\sqrt{P}\rho_{s}e^{-j2\pi\tau_{s}(n-1)\varDelta f}\mathbf{g}_{t}^{\mathsf{T}}\dot{\mathbf{a}}_{d_{s}}\left(\mathbf{p}_{s}\right),\\
\frac{\partial\mu_{t}[n]}{\partial\tau_{s}} & =-j2\pi\tau_{s}(n-1)\varDelta f\sqrt{P}\rho_{s}e^{-j2\pi\tau_{s}(n-1)\varDelta f}\mathbf{g}_{t}^{\mathsf{T}}\mathbf{a}\left(\mathbf{p}_{s}\right),
\end{align}
where $\dot{\mathbf{a}}_{x}(\mathbf{p}_{s})\triangleq\partial\mathbf{a}(\mathbf{p}_{s})/\partial x$.

\section{Derivation of the Jacobian Matrix}
According to the definition of $\mathbf{J}\triangleq\partial\boldsymbol{\eta}^{\mathsf{T}}/\partial\boldsymbol{\eta}_{p}$,
we can obtain the derivatives as follows:
\begin{align}
\frac{\partial\phi_{el,s}}{\partial x_{s}} & =\frac{-(x_{s}-x_{\mathrm{R}})(z_{s}-z_{\mathrm{R}})}{\sqrt{(x_{s}-x_{\mathrm{R}})^{2}+(y_{s}-y_{\mathrm{R}})^{2}}\left\Vert \mathbf{p}_{s}-\mathbf{p}_{\mathrm{R}}\right\Vert ^{2}},\\
\frac{\partial\phi_{el,s}}{\partial y_{s}} & =\frac{-(y_{s}-y_{\mathrm{R}})(z_{s}-z_{\mathrm{R}})}{\sqrt{(x_{s}-x_{\mathrm{R}})^{2}+(y_{s}-y_{\mathrm{R}})^{2}}\left\Vert \mathbf{p}_{s}-\mathbf{p}_{\mathrm{R}}\right\Vert ^{2}},\\
\frac{\partial\phi_{el,s}}{\partial z_{s}} & =\frac{\sqrt{(x_{s}-x_{\mathrm{R}})^{2}+(y_{s}-y_{\mathrm{R}})^{2}}}{\left\Vert \mathbf{p}_{s}-\mathbf{p}_{\mathrm{R}}\right\Vert ^{2}},
\end{align}
\begin{flalign}
\frac{\partial\phi_{az,s}}{\partial x_{s}} & =\frac{-\mathrm{sgn}(x_{s}-x_{\mathrm{R}})(y_{s}-y_{\mathrm{R}})}{(x_{s}-x_{\mathrm{R}})^{2}+(y_{s}-y_{\mathrm{R}})^{2}},\\
\frac{\partial\phi_{az,s}}{\partial y_{s}} & =\frac{-\mathrm{sgn}(x_{s}-x_{\mathrm{R}})(x_{s}-x_{\mathrm{R}})}{(x_{s}-x_{\mathrm{R}})^{2}+(y_{s}-y_{\mathrm{R}})^{2}},
\end{flalign}
\begin{align}
\frac{\partial d_{s}}{\partial x_{s}} & =\frac{x_{s}-x_{\mathrm{R}}}{\left\Vert \mathbf{p}_{s}-\mathbf{p}_{\mathrm{R}}\right\Vert },\\
\frac{\partial d_{s}}{\partial y_{s}} & =\frac{y_{s}-y_{\mathrm{R}}}{\left\Vert \mathbf{p}_{s}-\mathbf{p}_{\mathrm{R}}\right\Vert },\\
\frac{\partial d_{s}}{\partial z_{s}} & =\frac{z_{s}-z_{\mathrm{R}}}{\left\Vert \mathbf{p}_{s}-\mathbf{p}_{\mathrm{R}}\right\Vert },
\end{align}
for $s\in\{0,1,...,N_{s}\}$. And
\begin{alignat}{2}
\frac{\partial\tau_{0}}{\partial x_{0}} & =\frac{x_{0}-x_{\mathrm{R}}}{c\left\Vert \mathbf{p}_{0}-\mathbf{p}_{\mathrm{R}}\right\Vert }, &\quad& \frac{\partial\tau_{s}}{\partial x_{0}} =\frac{x_{0}-x_{s}}{c\left\Vert \mathbf{p}_{0}-\mathbf{p}_{s}\right\Vert }, \\
\frac{\partial\tau_{0}}{\partial y_{0}} & =\frac{y_{0}-y_{\mathrm{R}}}{c\left\Vert \mathbf{p}_{0}-\mathbf{p}_{\mathrm{R}}\right\Vert }, &\quad& \frac{\partial\tau_{s}}{\partial y_{0}} =\frac{y_{0}-y_{s}}{c\left\Vert \mathbf{p}_{0}-\mathbf{p}_{s}\right\Vert }, \\
\frac{\partial\tau_{0}}{\partial z_{0}} & =\frac{z_{0}-z_{\mathrm{R}}}{c\left\Vert \mathbf{p}_{0}-\mathbf{p}_{\mathrm{R}}\right\Vert }, &\quad& \frac{\partial\tau_{s}}{\partial z_{0}} =\frac{z_{0}-z_{s}}{c\left\Vert \mathbf{p}_{0}-\mathbf{p}_{s}\right\Vert },
\end{alignat}
\begin{align}
\frac{\partial\tau_{s}}{\partial x_{s}} & =\frac{x_{s}-x_{\mathrm{R}}}{c\left\Vert \mathbf{p}_{s}-\mathbf{p}_{\mathrm{R}}\right\Vert }+\frac{x_{s}-x_{0}}{c\left\Vert \mathbf{p}_{0}-\mathbf{p}_{s}\right\Vert },\\
\frac{\partial\tau_{s}}{\partial y_{s}} & =\frac{y_{s}-y_{\mathrm{R}}}{c\left\Vert \mathbf{p}_{s}-\mathbf{p}_{\mathrm{R}}\right\Vert }+\frac{y_{s}-y_{0}}{c\left\Vert \mathbf{p}_{0}-\mathbf{p}_{s}\right\Vert },\\
\frac{\partial\tau_{s}}{\partial z_{s}} & =\frac{z_{s}-z_{\mathrm{R}}}{c\left\Vert \mathbf{p}_{s}-\mathbf{p}_{\mathrm{R}}\right\Vert }+\frac{z_{s}-z_{0}}{c\left\Vert \mathbf{p}_{0}-\mathbf{p}_{s}\right\Vert },
\end{align}
for $s\in\{1,...,N_{s}\}$. The remaining elements in the Jacobian matrix $\mathbf{J}$ are zero.

\section{Derivation of \eqref{eq:60}}
We can obtain the derivatives as follows:
\begin{align}
\frac{\partial\boldsymbol{\kappa}[n]}{\partial\mathfrak{R}(\rho_{s})} & =\sqrt{P}e^{-j2\pi\tau_{s}(n-1)\varDelta f}\mathbf{b}\left(\mathbf{p}_{s}\right),\\
\frac{\partial\boldsymbol{\kappa}[n]}{\partial\mathfrak{I}(\rho_{s})} & =j\sqrt{P}e^{-j2\pi\tau_{s}(n-1)\varDelta f}\mathbf{b}\left(\mathbf{p}_{s}\right),\\
\frac{\partial\boldsymbol{\kappa}[n]}{\partial\phi_{el,s}} & =\sqrt{P}\rho_{s}e^{-j2\pi\tau_{s}(n-1)\varDelta f}\dot{\mathbf{b}}_{\phi_{el,s}}\left(\mathbf{p}_{s}\right),\\
\frac{\partial\boldsymbol{\kappa}[n]}{\partial\phi_{az,s}} & =\sqrt{P}\rho_{s}e^{-j2\pi\tau_{s}(n-1)\varDelta f}\dot{\mathbf{b}}_{\phi_{az,s}}\left(\mathbf{p}_{s}\right),\\
\frac{\partial\boldsymbol{\kappa}[n]}{\partial d_{s}} & =\sqrt{P}\rho_{s}e^{-j2\pi\tau_{s}(n-1)\varDelta f}\dot{\mathbf{b}}_{d_{s}}\left(\mathbf{p}_{s}\right),
\end{align}
\begin{align}
\frac{\partial\boldsymbol{\kappa}[n]}{\partial d_{s}} & =-j\sqrt{P}2\pi\tau_{s}(n-1)\varDelta f\rho_{s}e^{-j2\pi\tau_{s}(n-1)\varDelta f}\mathbf{b}\left(\mathbf{p}_{s}\right),
\end{align}
where $\dot{\mathbf{b}}_{x}(\mathbf{p}_{s})\triangleq\partial\mathbf{b}(\mathbf{p}_{s})/\partial x$.

And $\mathbf{K}_{s}[n]$ can be represented as
\begin{align}
\mathbf{K}_{s}[n] & =[\mathbf{b}\left(\mathbf{p}_{s}\right),\dot{\mathbf{b}}_{\phi_{el,s}}\left(\mathbf{p}_{s}\right),\dot{\mathbf{b}}_{\phi_{az,s}}\left(\mathbf{p}_{s}\right),\dot{\mathbf{b}}_{d_{s}}\left(\mathbf{p}_{s}\right)]\nonumber \\
\times\sqrt{P} & \begin{bmatrix}
e^{\varsigma_{n}} & je^{\varsigma_{n}} & 0 & 0 & 0 & -j\varsigma_{n}\rho_{s}e^{\varsigma_{n}}\\
0 & 0 & \rho_{s}e^{\varsigma_{n}} & 0 & 0 & 0\\
0 & 0 & 0 & \rho_{s}e^{\varsigma_{n}} & 0 & 0\\
0 & 0 & 0 & 0 & \rho_{s}e^{\varsigma_{n}} & 0
\end{bmatrix},\label{eq:98}
\end{align}
where $\varsigma_{n}\triangleq-j2\pi\tau_{s}(n-1)\varDelta f$.
}

\bibliographystyle{IEEEtran}
\bibliography{reference}

\end{document}